\documentclass[12pt,a4paper,twoside]{article}
\usepackage{epsfig}
\newlength{\figwidth}
\newlength{\figheight}
\newlength{\twofigheight}
\def\gg{$\gamma \gamma$}

\def\wgg{$W_{\gamma \gamma}$}

\def\ggg{$\Gamma_{\gamma \gamma}$}

\def\pgg{$\phi_{\gamma \gamma}$}

\def\epem{$e^+ e^-$}

\def\z0{Z}

\title{
       \vspace{-1.5cm}
       \begin{center}
 {\small    European Physical Society} \\[-3mm]
{\small International Europhysics Conference on High Energy Physics} \\[-3mm]
{\small EPS (July 17th-23rd 2003) in Aachen, Germany} 
       \end{center}
       \vspace{1cm}
       \begin{flushright}
{\normalsize\bf    Abstract ID: 605}
       \end{flushright}
       \vspace{1cm}
 Measurement of the Higgs-boson CP properties using decays into 
WW and ZZ at the Photon Collider
      }
 
\author{\bf Photon Collider Physics Working Group (ECFA Study) \\[5mm]
{\small P.Nie\.zurawski, A.F.\.Zarnecki} \\
{\small\it Institute of Experimental Physics, Warsaw University,} \\
{\small M.Krawczyk} \\
{\small\it Institute of Theoretical Physics, Warsaw University} 
} 

\date{}

\begin{document} 

\maketitle 

\vfill

\begin{abstract}
Higgs-boson production at the Photon Collider at
TESLA is studied for masses from 180 to 350~GeV,
using realistic luminosity spectra and detector simulation. 
Parity of the SM Higgs-boson can be verified 
from the measurement of 
the interference effects in the $W^+ W^-$ decay channel
and of the angular correlations in the decays of $W^+ W^-$ and 
$Z Z$ pairs.
SM-like Two Higgs Doublet Model (2HDM~II) and
model with the generic higgs couplings are considered.
\end{abstract}

\vfill


\thispagestyle{empty}

\clearpage

%
%

\section{Introduction}
\label{sec:intro}

A photon collider has been  proposed as a natural
extension of the \epem\ linear collider project TESLA \cite{tdr_pc}.
The physics potential of a photon collider is very
rich and complementary to the physics program of the \epem\
and hadron-hadron colliders.
It is an ideal place to study the mechanism of the electroweak 
symmetry breaking (EWSB) and the properties of the Higgs-boson.
In paper \cite{nzk_wwzz} we performed realistic simulation of 
SM Higgs-boson production at the Photon Collider
for $W^+ W^-$ and $\z0 \z0$ decay channels,
for Higgs-boson masses above 150~GeV.
Due to an interference with a large  Standard Model background, 
the process $\gamma \gamma \rightarrow higgs \rightarrow W^+ W^- / \z0 \z0$
turns out to be sensitive not only to the 
Higgs-boson partial width to \gg, \ggg, but also to 
the phase of the $\gamma \gamma \rightarrow h $ coupling, \pgg.
A precise measurement of both \ggg\ and \pgg\ 
seems to be crucial for determination of the Higgs-boson couplings,
see also \cite{ginzburg,cros_zz,gounaris,asakawa,hagiwara}.
In \cite{nzk_wwzz} we have found that this is extremely important
to combine both  $W^+ W^-$ and $ZZ$ channels, as the first one due to a 
large background is is very sensitive to a phase, while the second - 
to a partial width.
From combined analysis of  $W^+ W^-$ and $\z0 \z0$ decay channels
the $\gamma \gamma$ partial width can be measured with an accuracy of 3 to 8\%
and the  phase of the amplitude with an accuracy between 30 and 100~mrad.

In this paper we extend this analysis 
by considering in detail various  extensions of the Standard Model,
in particular the SM-like Two Higgs Doublet Model 2HDM~II with and 
without CP-conservation.
%
%
%
We perform the combined analysis of $W^+ W^-$ and $ZZ$ invariant-mass 
distributions and extract the corresponding Higgs-boson couplings 
($\tan\beta$).
For 2HDM model with a CP violation, we estimate precision of the $H-A$ mixing 
angle. 

We also consider here a potential of establishing CP-properties
of Higgs-bosons in a general CP-violating model  from  a
combined analysis of the angular distributions 
of the  $W^+ W^-$ and $ZZ$ decay-products. 
In this approach a CP properties of the Higgs boson can be studied
independently on a production mechanism.
We consider a  model with  generic, CP-violating higgs couplings 
to vector bosons, 
%
leading to   different  angular distributions for a scalar- and 
pseudoscalar-type of  couplings.  From such measurement 
 the CP-parity of the observed Higgs state can be determined.
Precision in the determination of couplings can be significantly
improved if measurement of  angular distributions is combined with
the  invariant-mass one.
However, 
in such case  additional model assumptions are needed.

%
%

\section{Measurement of the $h(H) \rightarrow \gamma \gamma$ 
  width and phase \\ from the invariant-mass distributions }
\label{sec:mdist}

%


The Compton back-scattering of a laser light off  high-energy
electron beams is considered as a  source of high energy, highly polarized
photon beams~\cite{pc}. 
According to the  current design \cite{tdr_pc}, the energy
of the laser photons is assumed to be fixed
for all considered electron-beam energies.
%
%
With 100\% circular polarization of laser photons 
and 85\% longitudinal polarization of the electron beam
the luminosity spectra peaked at high $\gamma \gamma$ invariant masses
is expected.

The analysis  bases on the CompAZ parametrization \cite{compaz} of
the realistic luminosity spectra for a Photon Collider 
at  TESLA \cite{telnov}.
We assume that the centre-of-mass energy of colliding electron beams 
$\sqrt{s_{ee}}$, is always optimised for the production of
a Higgs boson with a given mass.
The event generation 
according  to the cross-section formula 
for a vector-boson production including the higgs  contribution 
\cite{ginzburg,cros_zz,cros_ww}
is done with PYTHIA~6.152 \cite{PYTHIA}.
The fast simulation program SIMDET version 3.01 \cite{SIMDET}
is used to model the TESLA detector performance.
The selection cuts are  applied to select 
$W^+ W^- \! \rightarrow q\bar{q} q\bar{q}$
and $\z0  \z0  \rightarrow l\bar{l}q\bar{q}$ events
($l=\mu,\; e$).

All results presented in this paper were obtained 
for an integrated luminosity corresponding to one year 
of the photon collider running, as given by \cite{telnov}.
The total photon-photon luminosity increases from about
600~fb$^{-1}$ for $\sqrt{s_{ee}}=$305~GeV 
(optimal beam energy choice for $M=$200~GeV) to
about 1000~fb$^{-1}$ for $\sqrt{s_{ee}}=$500~GeV
(optimal beam energy choice for $M=$350~GeV).

\subsection{Standard Model analysis}
\label{sec:width}

In this section we summarise results of \cite{nzk_wwzz},
where the feasibility of measuring  Standard Model Higgs-boson production
in $W^+ W^-$ and $\z0 \z0$ decay channels  in \gg\ option of TESLA
has been studied for a Higgs-boson mass above 150~GeV.
We have studied the signal, i.e. the Higgs-boson decays into the 
vector bosons, and the background from direct vector-bosons production.
For the $\z0  \z0 $ final-state a direct, i.e. non-resonant 
$\gamma \gamma \rightarrow \z0 \z0$ process, is rare as it occurs   
via  loop only.  On contrary, the non-resonant $W^+W^-$ 
production is a tree-level process, and is expected to be large.
Therefore, also an interference between the signal of $W^+W^-$ production 
via the Higgs resonance and the background from direct production
may be large.
This effect can be used 
to access an information about the phase of
the $higgs \rightarrow \gamma \gamma $  coupling, \pgg.
For the Higgs-boson masses around 350~GeV we found that
the amplitude phase \pgg\ is more sensitive 
to the loop contributions of new,
heavy charged particles than  the \ggg\ itself.

\begin{figure}[p]
  \begin{center}
     \epsfig{figure=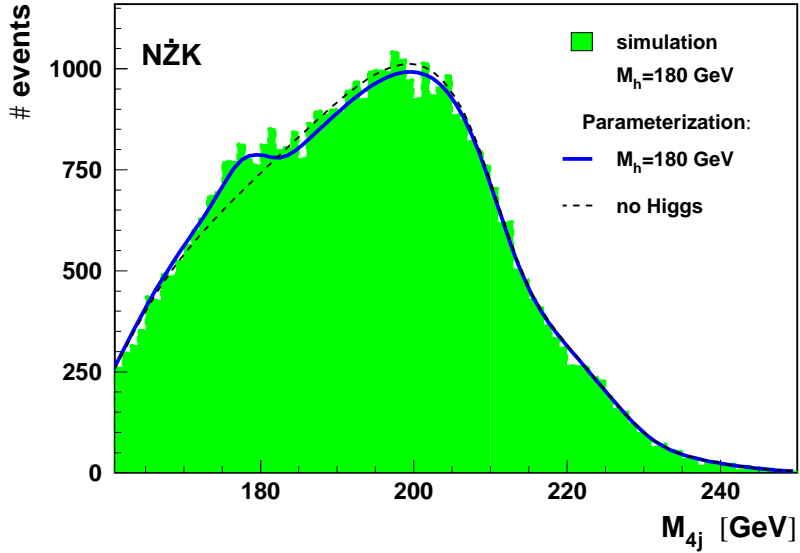,height=\twofigheight,clip=}
     \epsfig{figure=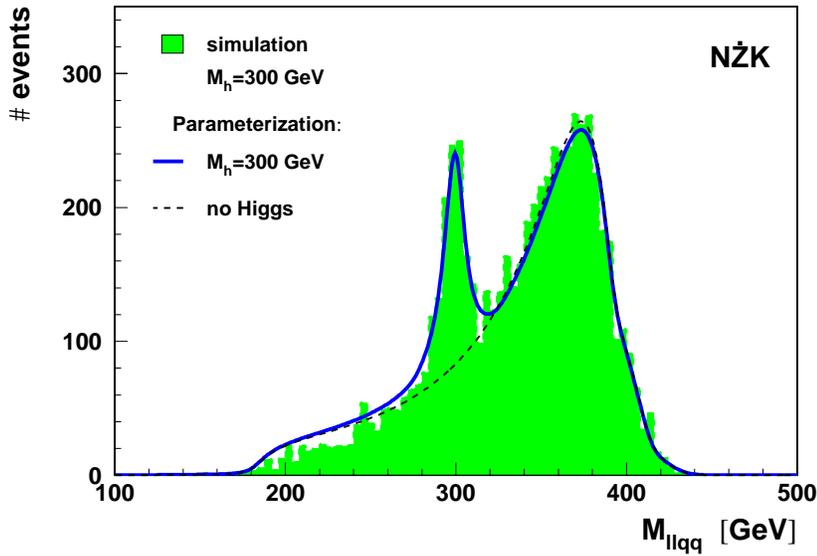,height=\twofigheight,clip=}
  \end{center}
  \caption{Distribution of the reconstructed invariant mass
        for $\gamma \gamma \rightarrow W^+ W^-$ events with a 
        SM Higgs-boson mass of 180~GeV and an 
        electron-beam energy of 152.5 GeV (upper plot) 
        and
        for $\gamma \gamma \rightarrow \z0  \z0 $ events, with a 
        SM Higgs-boson mass of 300~GeV and an 
        electron-beam energy of 250~GeV (lower plot). 
        Results from the simulation  based  on PYTHIA 
        and on the SIMDET detector-simulation (histogram) are 
        compared with the distribution
        obtained  by the numerical convolution of the cross-section 
        formula with the CompAZ  energy-spectra for photon-beams
        and parametrization of the detector resolution (solid line).
        The distribution expected without the higgs contribution is
        also shown (dashed line) \cite{nzk_wwzz}.
         } 
 \label{fig:convol} 
 \end{figure}

The invariant-mass resolution obtained from a  full simulation of $W^+ W^-$  
and $\z0 \z0$ events based on the PYTHIA and SIMDET programs, 
has been parametrized as a function of the \gg\ centre-of-mass energy, \wgg .
The distributions 
of the reconstructed invariant mass for 
$\gamma \gamma \rightarrow W^+ W^-$ 
 and $\gamma \gamma \rightarrow \z0  \z0 $ events
are shown in  Figs.~\ref{fig:convol}, upper and lower panels, respectively.
 Results from a full event simulation  based  on PYTHIA
 and SIMDET  are  compared with the distribution
 obtained from the numerical convolution of the relevant cross-sections
 with the CompAZ spectra and the parametrized detector resolution.


\begin{figure}[p]
  \begin{center}
     \epsfig{figure=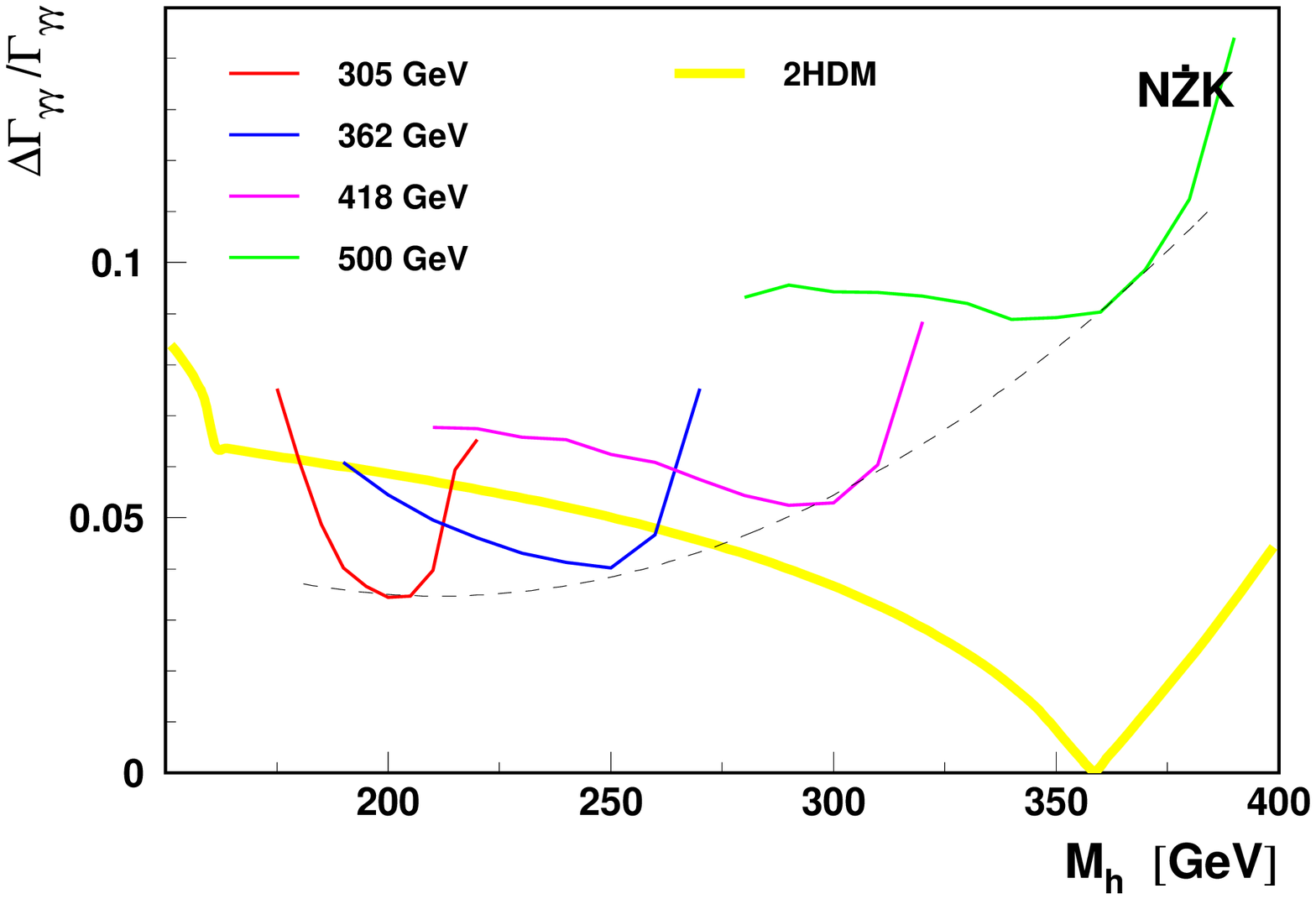,height=\twofigheight,clip=}
     \epsfig{figure=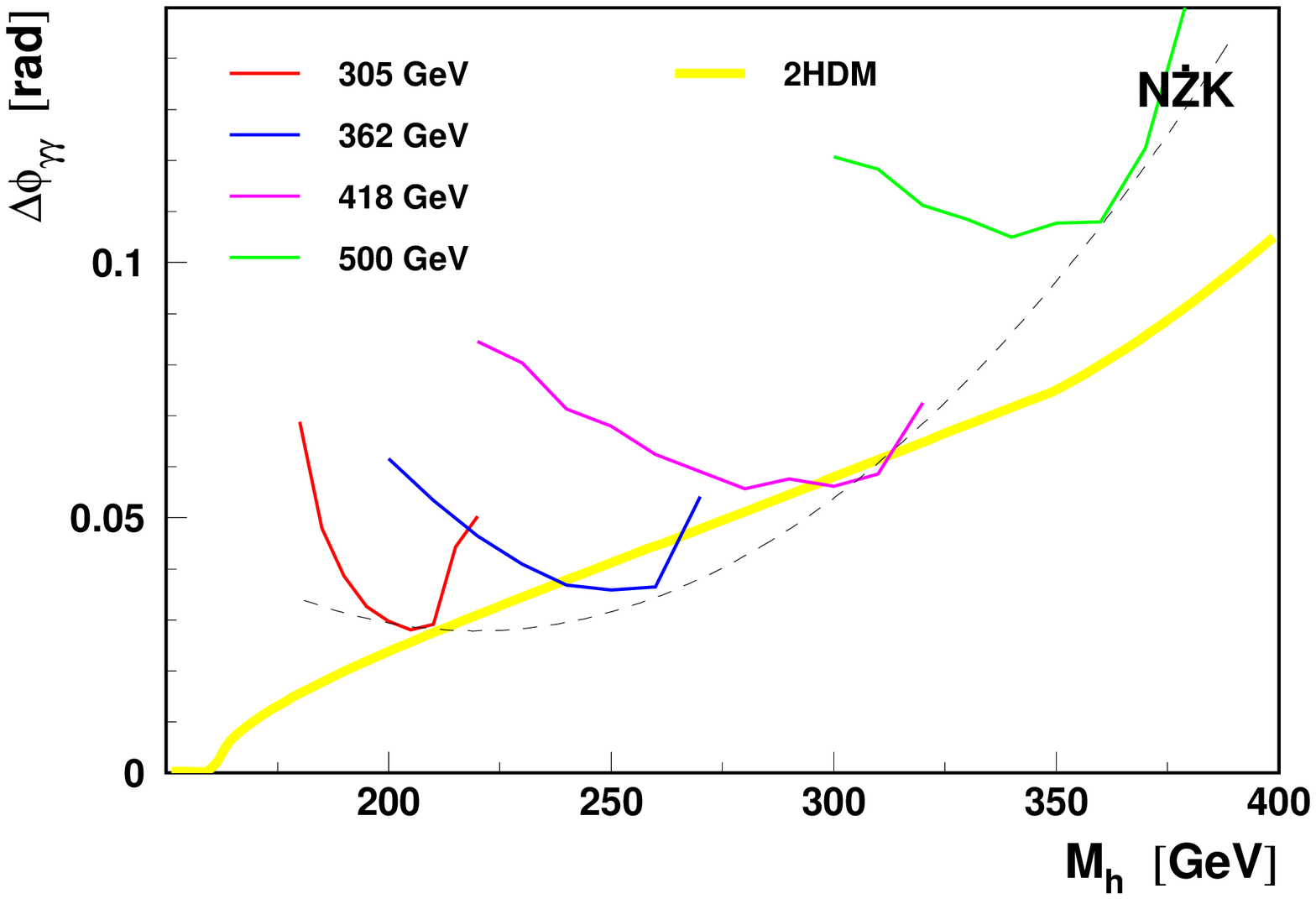,height=\twofigheight,clip=}
  \end{center}
  \caption{Statistical error in the 
           determination of the  Higgs-boson width \ggg\ (upper plot)
           and phase  \pgg\ (lower plot),
        from the combined fit to the observed $W^+ W^-$ and $ZZ$ mass spectra,
          as a function  of the Higgs-boson mass $M_h$.
          Results are given for four various centre-of-mass
          energies of colliding electron beams $\sqrt{s_{ee}}$, as indicated
          in the plot.
          The yellow (tick) line shows the size of 
          the deviations expected in the SM-like  2HDM~II ~\cite{2HDM}
          with an additional contribution due to the charged
          Higgs-boson of  mass $M_{H^+}=800$ GeV. 
          The thin dashed line is included to guide the eye. \cite{nzk_wwzz}
          }
 \label{fig:smfinal} 
 \end{figure}

Based on a parametric description of the expected mass distributions, 
many  experiments were simulated, each corresponding to one year
of a Photon Collider running at TESLA at a nominal luminosity.
The ``theoretical'' distributions were then fitted, simultaneously to
the observed $W^+ W^-$ and $\z0 \z0$ mass spectra, with the 
width \ggg\ and phase \pgg\ considered as the only free parameters.
Results of the fits performed for different Higgs-boson masses
and at different electron-beam energies are shown in Figs.~\ref{fig:smfinal}.
They indicate that with a proper choice of the electron-beam energy, the $\gamma \gamma$
partial width can be measured with an accuracy of 3 to 8\%, while
the  phase of the amplitude with an accuracy between 35 and 100~mrad,
see Fig. \ref{fig:smfinal}.
The \pgg\ measurement opens a new window to a precise
determination of the Higgs-boson couplings
and to  search of a ``new physics''.
It turns out that the phase is constrained predominantly by the $W^+ W^-$
invariant-mass distribution, thanks to large interference effects between
Higgs-boson decay and nonresonant $W^+ W^-$ production.
However, two-photon width of the Higgs-boson is much better constrained
by the measurement of the $\z0 \z0$ mass spectra, as the nonresonant
background is much smaller here.
A precise determination of both parameters is only possible
when both measurements are combined.


%


\subsection{Determination of the Higgs-boson couplings \\
        in the SM-like Two Higgs Doublet Model}
\label{sec:2hdm}

In our previous analysis \cite{nzk_wwzz}
we have considered possible deviations from the Standard Model predictions
resulting from the loop contributions to the $h \gamma \gamma$ vertex
 of new heavy charged particles. 
However, deviations in the two-photon width and phase
can also appear if the couplings of the Higgs-boson to the other particles are 
different than those predicted by the Standard Model.
This possibility is studied  in this paper, see also \cite{2HDM}.

As the simplest model with non-standard higgs couplings we consider 
Standard Model-like Two Higgs Doublet Model (2HDM~II) \cite{2HDM}.
We study a simple version of the so called solution B, 
where  the lightest Higgs-boson $h$ 
couplings to fermions are fixed and they
have the same values as in the Standard Model,
except that for up-type fermions the sign is opposite. The couplings of $h$
to the EW gauge-bosons may differ from the corresponding SM predictions.
There is only one parameter in such model, $\beta$,  and
all remaining couplings of $h$ and all couplings of $H$ and $A$
can be  expressed in terms of it, as presented in the 
table \ref{tab:2hdm}.\footnote{%
$h$ and $H$ couplings to the charged higgs boson $H^\pm$ are calculated 
according to the 2HDM II potential \cite{2HDM} assuming $\mu$=0.
}
%
%

\renewcommand{\arraystretch}{1.3}
\begin{table}[b]
\begin{center}
\begin{tabular}{|l|c|c|c|}
\hline
    & $h$ & $H$ & $A$ \\ \hline
$\chi_u $  & $-1$ & $-\frac{1}{\tan\beta}$ &  
  $-i  \; \gamma_5  \; \frac{1}{\tan\beta} $ \\
$\chi_d $  & $+1$ & $-\tan\beta$ &  
  $-i  \; \gamma_5  \; \tan\beta $ \\
$\chi_V $  & $\cos ( 2 \beta)$ & $- \sin ( 2 \beta)$ &    $0$ \\
\hline
\end{tabular}
\end{center}
\caption{
Couplings of the neutral Higgs-bosons to fermions and vector bosons,
relative to the Standard Model couplings, for the considered solution B 
of the SM-like Two Higgs Doublet Model (2HDM~II).
}
\label{tab:2hdm}
\end{table}

One expects significant deviations from the Standard Model predictions 
for a light Higgs-boson $h$, both for the two-photon width and phase, 
for $\tan \beta \ll 1$. 
As compared to the SM there is a change of a relative sign of the 
top-quark and the $W$ 
contribution, therefore the two-photon width in 2HDM II (B)
is significantly larger than
in the Standard Model, where these two contributions partly cancel each other.
For $\tan \beta \sim 1$ the two-photon width decreases, due to the
suppressed $W$-loop contribution ($g_{hWW} \sim 0$).
Finally, for large values of $\tan \beta$ ($\cos( 2 \beta)\approx -1$)  
the two-photon width of the light Higgs-boson $h$ tend to be very close 
to the expectations of the Standard Model. The only difference
is due to the presence of the charged Higgs-boson in the loop.
The difference resulting from the opposite, as compared to SM, 
 sign of the down-type fermion contributions is very small and 
can be neglected.

\begin{figure}[tb]
  \begin{center}
     \epsfig{figure=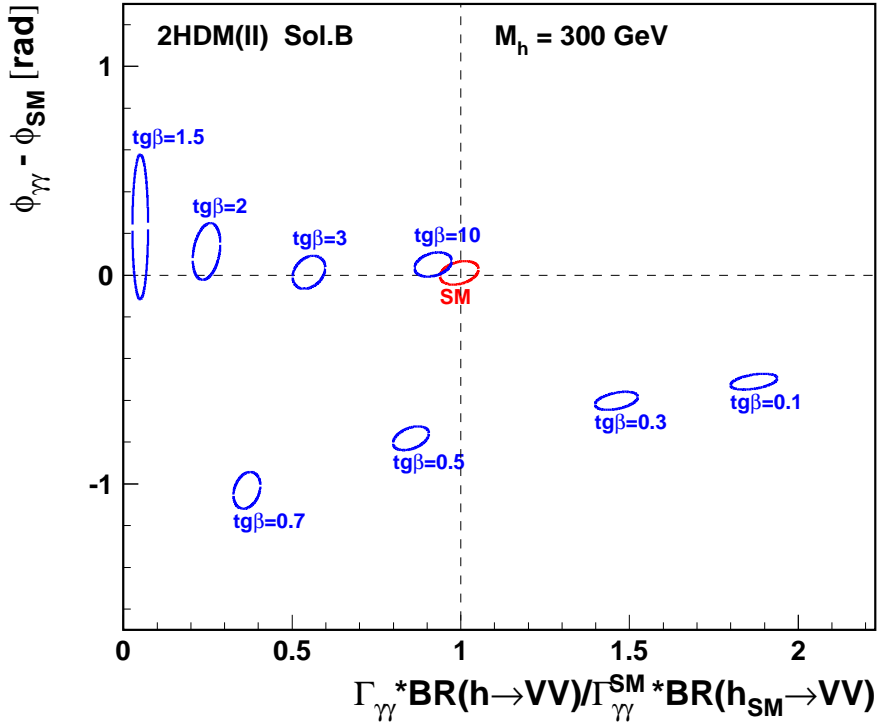,height=\figheight,clip=}
  \end{center}
 \caption{ 
         The deviation from the SM of
for the light Higgs-boson $h$ with mass 
$M_h = 300$ GeV in 
the SM-like 2HDM~II (B), 
with charged Higgs-boson mass of 800~GeV for different 
values of $\tan \beta$.
Error contours (1$\sigma$) on the measured deviation from
the Standard Model predictions for
the $h\rightarrow \gamma \gamma$ phase \pgg\ and for
the decay width \ggg\ times vector boson branching ratio $BR(h\rightarrow VV)$,
correspond to $L_{\gamma \gamma} \approx 840$~fb$^{-1}$. 
Red contour labelled 'SM' indicates the expected precision for the Standard 
Model.
 } 
 \label{fig:showel1} 
 \end{figure}

\begin{figure}[tb]
  \begin{center}
     \epsfig{figure=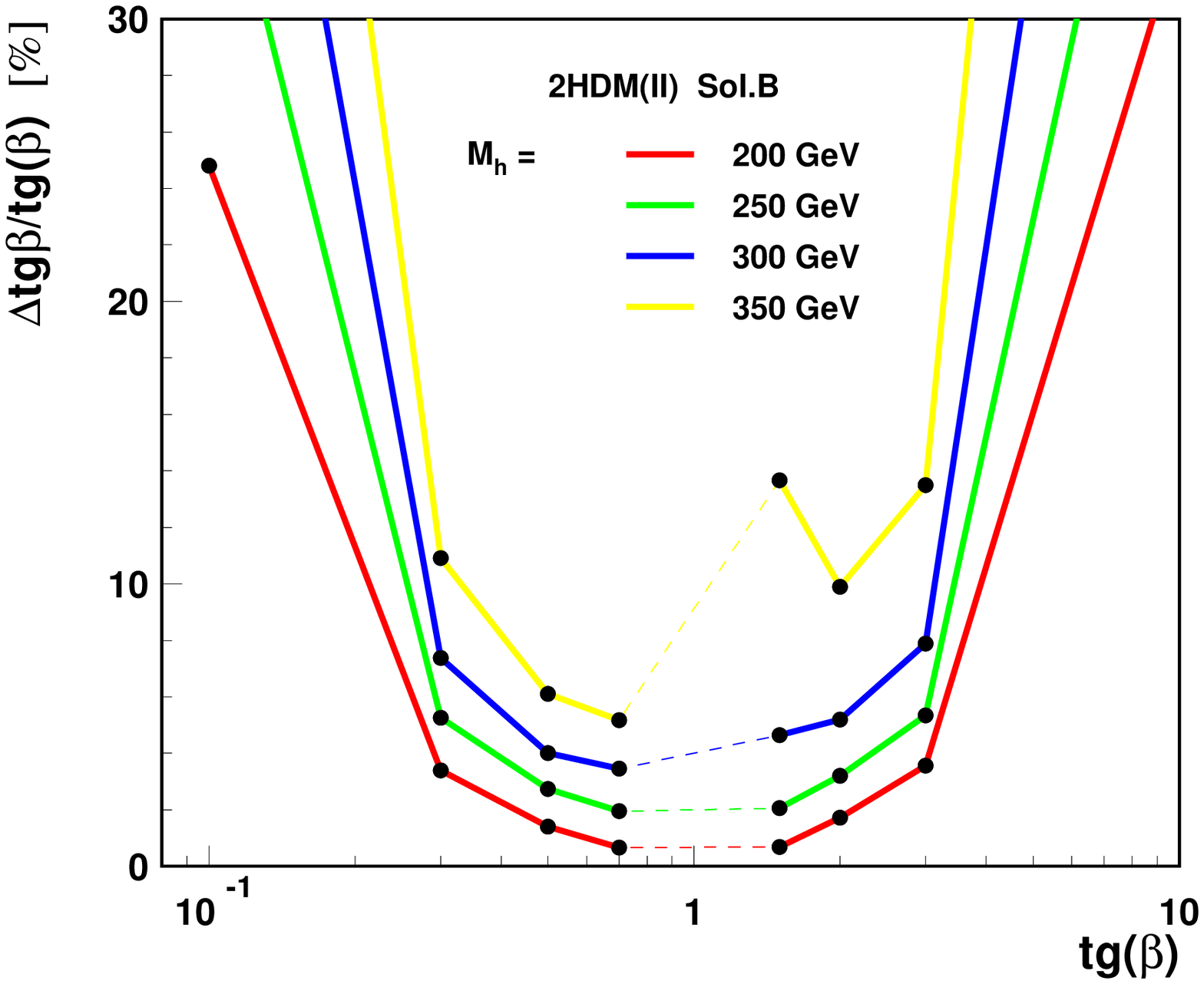,height=\figheight,clip=}
  \end{center}
 \caption{ 
          Statistical error in the determination of  $\tan \beta$,
          for four values of light Higgs-boson mass $M_h$. 
The simultaneous fit to the observed $W^+ W^-$ and  $ZZ$ mass spectra, 
is considered for the 
SM-like 2HDM~II (B), 
with charged Higgs-boson mass of 800~GeV.
Centre-of-mass energy of colliding electron beams $\sqrt{s_{ee}}$ 
is optimised for each mass $M_h$.
           } 
 \label{fig:errli} 
 \end{figure}

Results  for the 
light Higgs-boson $h$ with mass $M_h = 300$ GeV, 
 from the measurement of the two-photon 
width (times the vector-boson branching ratio) and phase 
are presented in Fig.~\ref{fig:showel1} for various $\tan \beta$ values.
Error contours (1$\sigma$) on the expected deviation from
the Standard-Model predictions correspond to one year of photon collider
running, i.e. $L_{\gamma \gamma} \approx 840$~fb$^{-1}$. 
In the combined fit to the invariant-mass distributions for 
$W^+ W^-$ and $Z Z$ events we take the advantage of the fact that the ratio
of the corresponding branching ratios 
$BR(h\rightarrow ZZ)/BR(h\rightarrow W^+ W^-)$ is expected to be the same
as in  the Standard Model.
Charged Higgs-boson mass is set to 800~GeV.
Results presented in Fig.~\ref{fig:showel1} show that the measurement
of the two-photon width and phase for the light Higgs-boson $h$
decaying into $W^+ W^-$ and $Z Z$ would allow  a precise determination 
of the $\tan \beta$ value. 
The possible ambiguity in the measurement of the two-photon width is
resolved by the phase measurement, which clearly distinguishes between
low $\tan \beta$ and large $\tan \beta$ solutions.
The statistical error on the extracted  $\tan \beta$ value
is shown in Fig.~\ref{fig:errli} for different values 
of light Higgs-boson mass $M_h$.

%
%
%

%

%
%
For all considered Higgs-boson masses 
the expected error in the $\tan \beta$ determination is smallest 
for $\tan \beta$ close to 1.
This is because the Higgs-boson coupling to the vector bosons
is most sensitive to  $\tan \beta$, for such  values.
The precision of $\tan \beta$ measurement, for  $\tan \beta \sim 1$,
deteriorates with an increase of the  Higgs-boson mass.
It changes from about 1\% for mass of 200~GeV to about 10\% for
mass of 350~GeV.
For very high and very low $\tan \beta$, when the
relative  Higgs-boson coupling to vector bosons is close to $\pm1$ 
(table \ref{tab:2hdm}),
precise measurement of $\tan \beta$ is not possible.

%
%
 
The measurement of the two-photon width and phase
has been  investigated also for the heavy scalar Higgs-boson $H$
of the SM-like Two Higgs Doublet Model (solution B), with couplings as 
given in table \ref{tab:2hdm}.
In this case we consider only $0.2 \le \tan \beta \le 1$,  since 
for  $\tan \beta > 1$ both the top-quark and $W$  
contributions are strongly suppressed 
and the precision of the measurement deteriorates fast.
For $\tan \beta \sim 1$ both the two-photon width and phase of the heavy 
scalar Higgs-boson $H$ are  close to the expectations of 
the Standard Model (for a given $M_H$). 
For decreasing values of $\tan \beta$  the $W$-loop contribution
decreases, while the top-quark one increases and
as a result, the two-photon width decreases slightly for 
$\tan \beta \sim 0.5$ and then starts to increase with $\tan \beta$.
Finally, for $\tan \beta \sim 0.1$ the  Higgs-boson decays to $c \bar{c}$
start to dominate. The expected number of events with the $W^+ W^-$ and $ZZ$
decays drops rapidly and the measurement becomes problematic again.

\begin{figure}[tb]
  \begin{center}
     \epsfig{figure=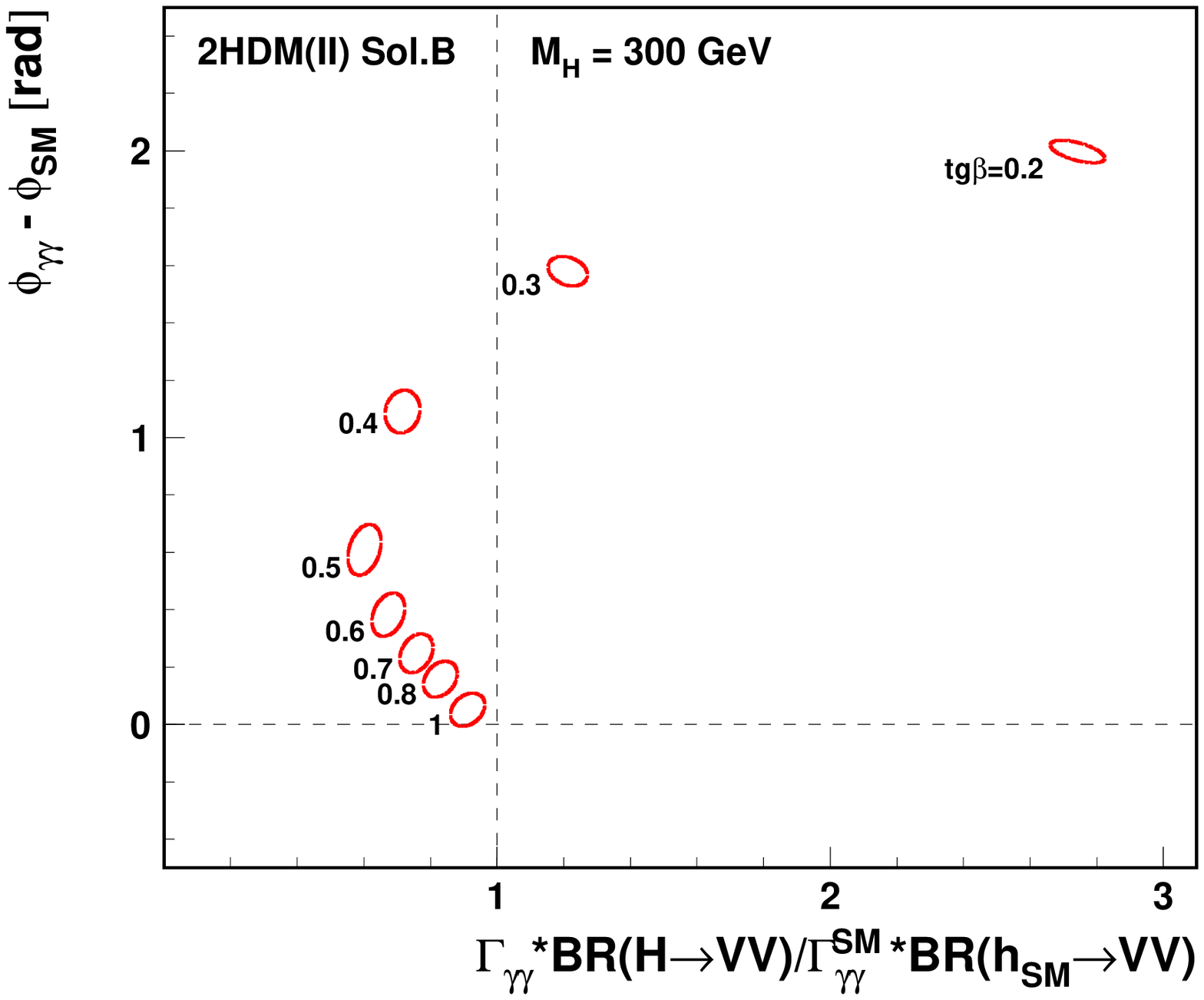,height=\figheight,clip=}
  \end{center}
 \caption{ 
     As in Fig. \ref{fig:showel1} for     the heavy
 Higgs-boson $H$ with mass $ 300$ GeV.
A light Higgs-boson mass is assumed to be $M_h = 120$ GeV.
         } 
 \label{fig:showel2} 
 \end{figure}

The results of the analysis of the measurement of the two-photon 
width  and phase, for the 
heavy scalar  $H$ with mass $M_h = 300$ GeV, 
are presented in Fig.~\ref{fig:showel2}.
Light Higgs-boson mass is set to 120~GeV and that of the 
charged Higgs-boson  to 800~GeV.

Error contours (1$\sigma$) on the measured deviation from
the Standard Model predictions are presented.
%
%
%
%
Results show that a precise determination of  $\tan \beta$, 
in the considered region of parameters, should also be possible 
for heavy scalar $H$ decaying to $W^+ W^-$ and $ZZ$.

%
%

\begin{figure}[tb]
  \begin{center}
     \epsfig{figure=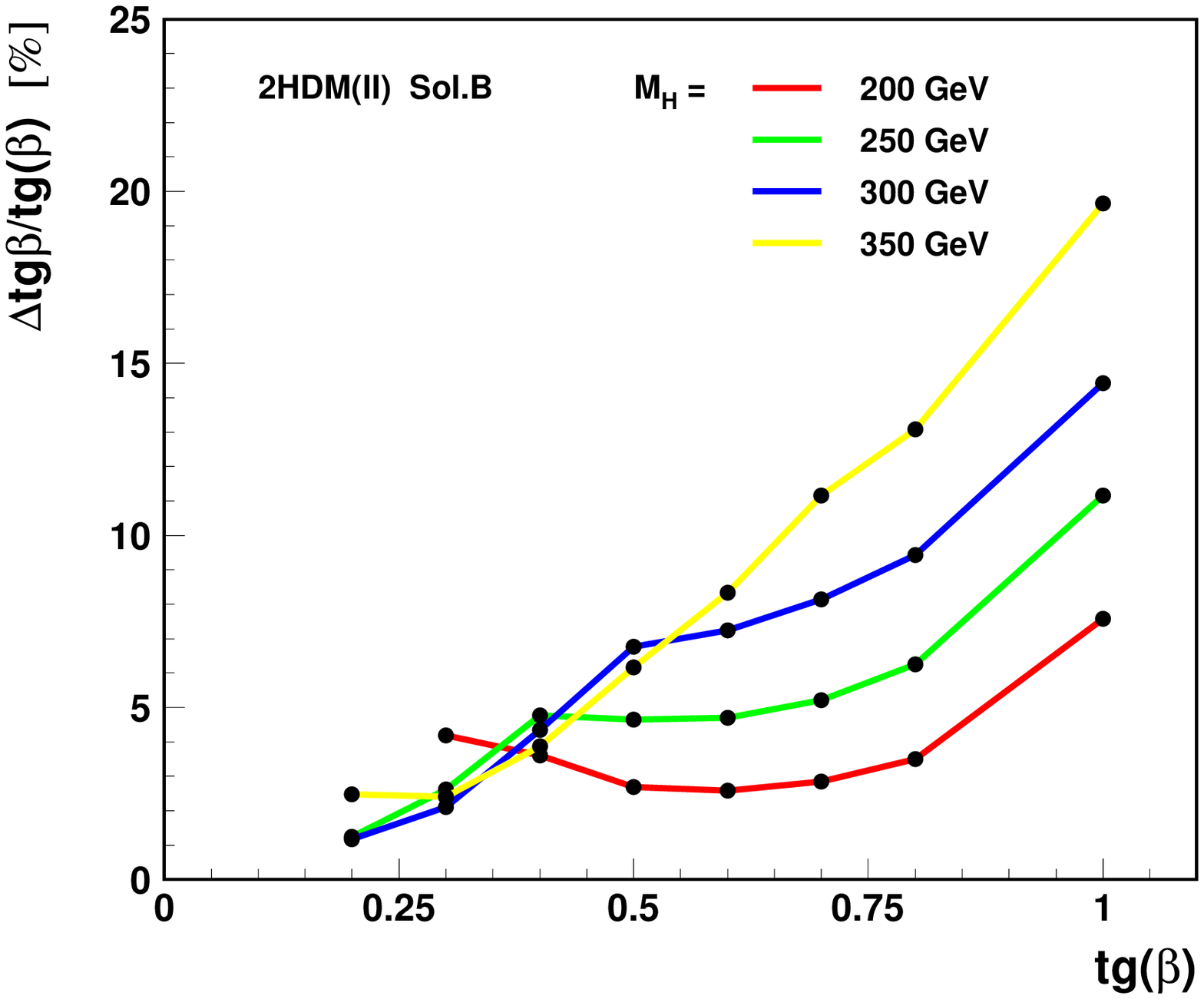,height=\figheight,clip=}
  \end{center}
 \caption{ 
As in Fig. \ref{fig:errli}
          for a  heavy Higgs boson $H$, 
with a light Higgs-boson mass of $M_h = 120$ GeV.
} 
 \label{fig:errhv} 
 \end{figure}

Shown in Fig.~\ref{fig:errhv} is the statistical error 
on the extracted  $\tan \beta$ values.
Results 
are given for four  values of heavy scalar Higgs-boson mass $M_H$,
 from 200 to 350 GeV.
The expected error in the $\tan \beta$ determination is small 
(1--3 \%) for  low  $\tan \beta$,
$\tan \beta \approx $0.25.
For larger values of $\tan \beta$,  the precision depends strongly on the
Higgs-boson mass. 
For mass of 200~GeV it changes between 3 and 8 \%,
whereas for mass of 350~GeV it is between 3 and 20 \%.

%
%

\subsection{Determination of CP properties of the Higgs boson \\
            in the SM-like Two Higgs Doublet Model}
\label{sec:cphiggs}

In the  general  Two Higgs Doublet Model \cite{CP2HDM},
the mass eigenstates of the neutral Higgs-bosons
$h_1$, $h_2$ and $h_3$
do not  match CP eigenstates $h$, $H$ and $A$.
We  consider here  a SM-like version of the CP-violating 
Two Higgs Doublet Model,  with CP violation
through a small mixing between $H$ and $A$ states.
%
%
We consider a simple version of such model, where
the couplings of the lightest higgs mass-eigenstate $h_1$ (with mass 120 GeV) 
are expected to correspond to the couplings of  $h$ boson,
whereas couplings of $h_2$ and $h_3$ states can be described
as the superposition of $H$ and $A$ couplings (see table \ref{tab:2hdm}), as follows:
\begin{eqnarray}
 \chi^{h_1}_X & \approx  & \chi^h_X  \nonumber \\*
 \chi^{h_2}_X & \approx  & 
        \chi^H_X \cdot \cos \Phi_{HA} \; + \; \chi^A_X \cdot \sin \Phi_{HA}
                                                       \label{EQ} \\*
 \chi^{h_3}_X & \approx  & 
      \chi^A_X \cdot \cos \Phi_{HA} \; - \; \chi^H_X  \cdot \sin \Phi_{HA}
                                                       \nonumber
\end{eqnarray}
where $X$ denotes a fermion or a vector boson, $X=u,\; d, \; V=W$ or $Z$.
We study the feasibility of the mixing angle $\Phi_{HA}$ determination
from the combined measurement of the two-photon width and phase for the 
higgs mass-eigenstate $h_2$.
We consider only small CP-violation effects, i.e. $|\Phi_{HA}| \ll 1$.

%
\begin{figure}[tb]
  \begin{center}
     \epsfig{figure=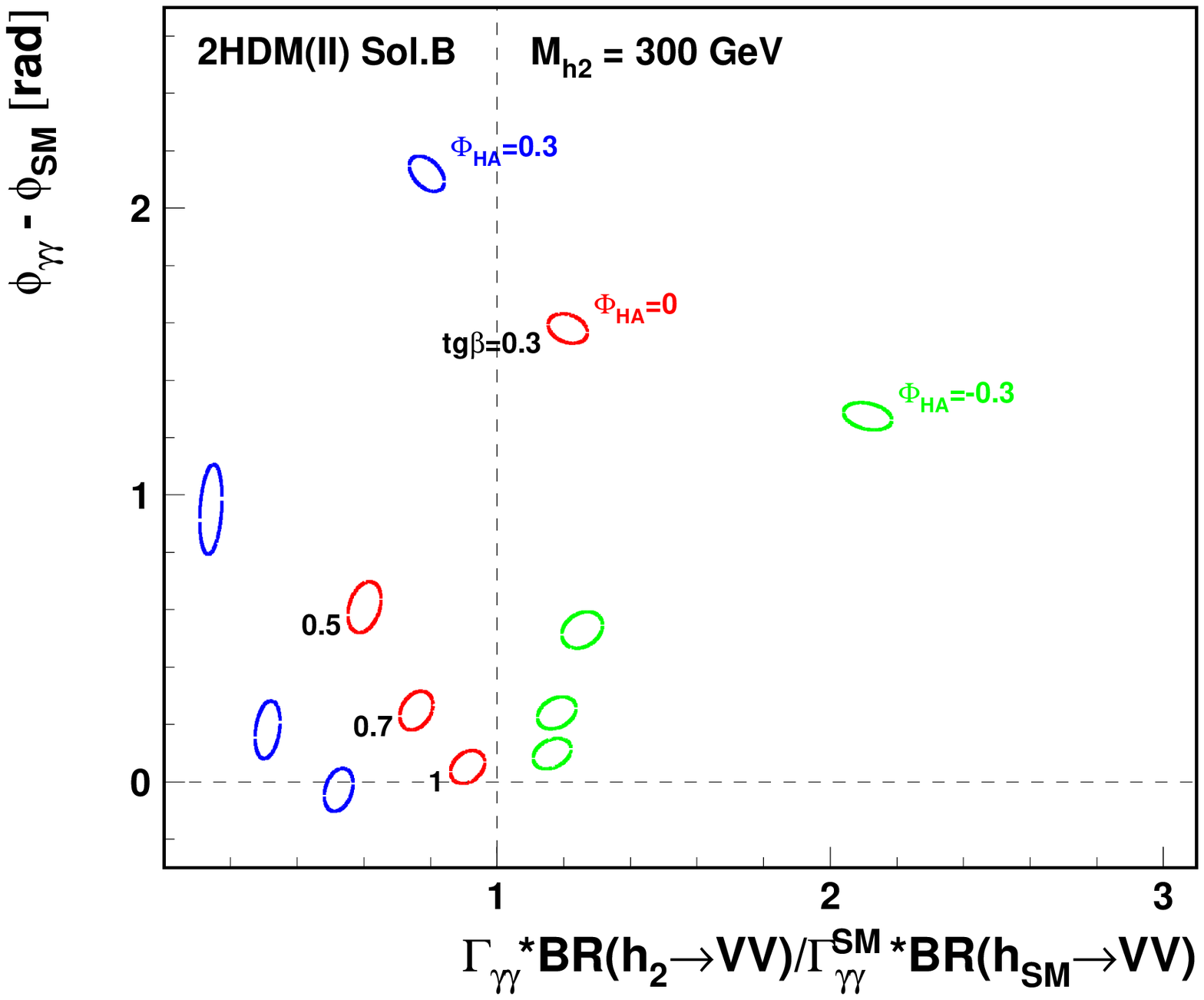,height=\figheight,clip=}
  \end{center}
 \caption{  
As in Fig.\ref{fig:showel2} for the SM-like 2HDM II (B) with CP-violation
for    the heavy
 Higgs-boson $h_2$ with mass $ 300$ GeV and couplings 
from Eq.~\ref{EQ}. 
A light Higgs-boson has mass  $M_{h_1} = 120$~GeV.
Three values of $H-A$ mixing angle
$\Phi_{HA}=-0.3,0,0.3$ are considered.
        } 
 \label{fig:showel3} 
 \end{figure} 
%

Results of the combined analysis (WW/ZZ decay channels)  
of  the  measurement of the two-photon 
width (times vector-boson branching ratio) and phase for the 
scalar Higgs-boson $h_2$ with mass $M_{h_2} = 300$ GeV, 
are presented in Fig.~\ref{fig:showel3}.
A light Higgs-boson has mass  $M_{h_1} = 120$~GeV while $M_{H^{\pm}}=800$ GeV.
The simultaneous fit to the observed $W^+ W^-$ and  $ZZ$ mass-spectra
was performed. 
Error contours (1$\sigma$) on the measured deviation from
the Standard Model predictions
are shown for    $\Phi_{HA} =0$, i.e. when   CP is conserved,
and for  CP violation with  $\Phi_{HA} = \pm 0.3$.
Even a small CP-violation can significantly influence  the
measured  two-photon width and two-photon phase, and therefore
it is possible to determine precisely  both the CP-violating  mixing 
angle $\Phi_{HA}$ and $\tan \beta$.

\begin{figure}[tb]
  \begin{center}
     \epsfig{figure=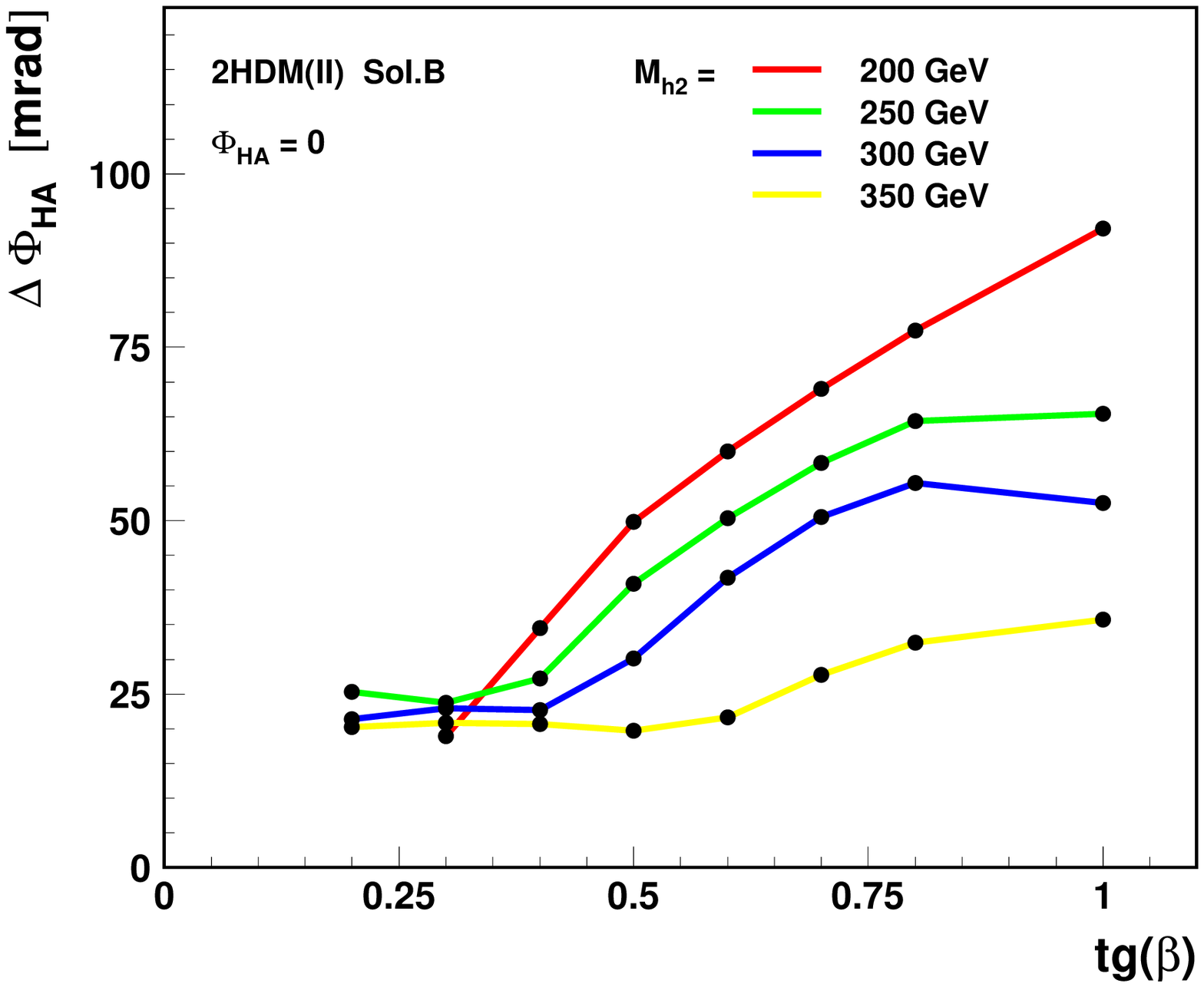,height=\figheight,clip=}
  \end{center}
 \caption{ 
    Verifying a   CP-conservation for the 
SM-like 2HDM~II (B).  Statistical error in the determination of 
the $H-A$ mixing angle $\Phi_{HA}$, 
as a function  of  $\tan \beta$ value, for four
values of heavy Higgs-boson mass $M_{h_2}$. 
The simultaneous fit to the observed $W^+ W^-$ and  $ZZ$ mass spectra, 
with light Higgs-boson mass of $120$ GeV,
charged Higgs-boson mass of 800~GeV, and no $H-A$ mixing ($\Phi_{HA}=0$),
Eq.~\ref{EQ}.
         } 
 \label{fig:errph} 
 \end{figure}

Next, we address a question: how well one can establish the fact
that in a model CP is conserved ? The answer can be read out 
from  Fig.~\ref{fig:errph} where the statistical error 
in the determination of the $H-A$ mixing angle
$\Phi_{HA}$, around $\Phi_{HA}=0$ value, is shown.
The results are presented  as a function of  $\tan \beta$ 
for four values of Higgs-boson mass $M_{h_2}$, from 200 to 350 GeV.
As above, we assume  a light Higgs-boson mass   $M_{h_1}=M_h = 120$ GeV,
charged Higgs-boson mass of 800~GeV.

%
For  low value of $\tan \beta$, $\tan \beta \approx$ 0.25,
the expected error is of the order of 20--25~mrad and does not
depend on the heavy Higgs-boson mass.
For larger values of $\tan \beta$  the precision worsens
for small Higgs-boson masses.
For $\tan \beta = 1$ the expected error changes 
from 35~mrad for mass of 350~GeV to about 90~mrad for mass of 200~GeV.

%
%

To summarize this part,
in the 2HDM model  the 
Higgs-boson couplings to the vector bosons are always   as for a scalar 
(say in the SM) with a modification by some functions, 
i.e. they are proportional to $g_h^{SM}$.
This is because there is no pseudoscalar coupling to $W^+ W^-$ and  $ZZ$
in this model ($\chi^A_V \equiv 0$).
The measured $W^+ W^-$ and  $ZZ$ mass-spectra are sensitive to the
CP-violating $H-A$ mixing angle only via a Yukawa coupling, more precisely
due of the top-quark loop
 contribution to the $h_2 \gamma \gamma $ vertex.
In the next section we will consider more general case, with a 
generic pseudoscalar couplings to the gauge bosons.


\section{Secondary decay analysis \\ for generic model with CP violation.}
\label{sec:cphiggs2}

In the general case, with a generic pseudoscalar coupling 
of Higgs-boson to  vector bosons,
not only the invariant-mass distribution of the reconstructed $W^+W^-$
and $ZZ$ events, as discussed in sec. 2.3, 
but also the angular distributions of $W^+W^-$
and $ZZ$ decay products are sensitive to the CP properties of the
Higgs-boson \cite{miller}. We use the following form of a coupling to 
vector boson $Z$ of a scalar ($H$) and a pseudoscalar ($A$):
\begin{eqnarray}
g_{HZZ}  =  i g \frac{M_Z}{\cos \theta_W} \; \left(
  \lambda_H \cdot g^{\mu \nu} \right)\ 
= g_{HZZ}^{SM} \;  \lambda_H \ , \\[3mm]
g_{AZZ} = i g \frac{M_Z}{\cos \theta_W} \; \left(
  \lambda_A \cdot \varepsilon^{\mu \nu \rho \sigma}\; 
  \frac{(p_1+p_2)_\rho \; (p_1-p_2)_\sigma}{M_Z^2} \right)\ , 
\label{ha}
\end{eqnarray}
where $p_1$ and $p_2$ are the 4-momenta of the vector bosons.

The coupling with $\lambda_H$  corresponds to the SM-like 
CP-even coupling, whereas the one  with  $\lambda_A$ is
a general CP-odd coupling for the spin-0 boson (higgs).
And we use similar expressions for the coupling to $W$, 
with an obvious change $M_Z$ to $M_W$ in the denominator of the 
pseudoscalar coupling, and a change in normalization factor
as in the SM, namely $M_Z/\cos \theta_W \rightarrow M_W$.
We see, that if $\lambda_H=1$, the scalar $H$ behaves as the SM Higgs boson.

\subsection{Angular distributions for secondary decays}

\begin{figure}[b]
  \begin{center}
     \epsfig{figure=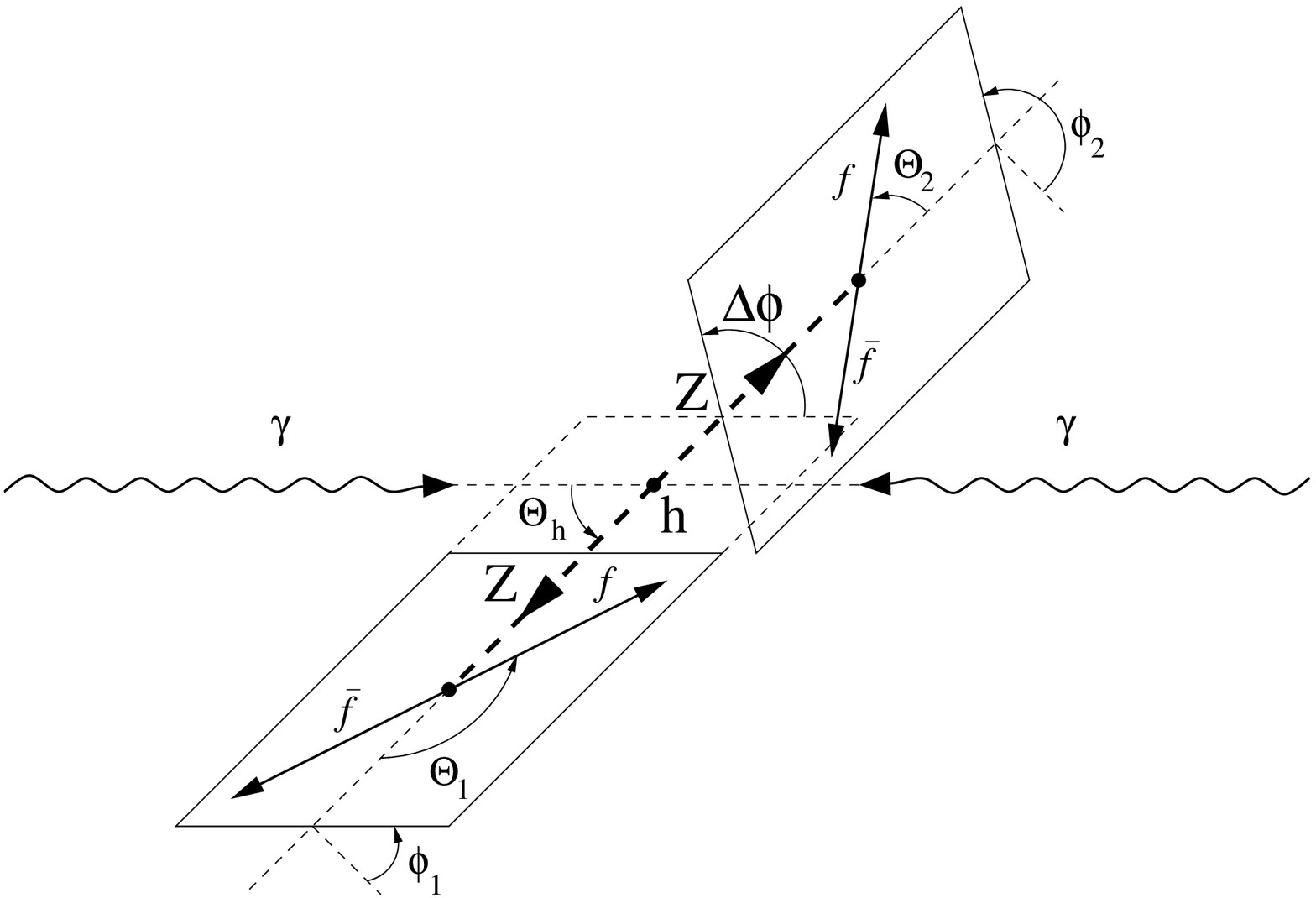,height=\figheight,clip=}
  \end{center}
 \caption{
    The definition of the polar angles $\Theta_h$, $\Theta_1$ and $\Theta_2$,
  and the azimuthal angles $\phi_1$ and $\phi_2$ for the process
$\gamma \gamma \rightarrow h \rightarrow Z Z \rightarrow 4\; f $.
$\Delta \phi$ is the angle between two Z decay planes,
$\Delta \phi = \phi_2 - \phi_1$. 
All polar angles are calculated in the rest frame of the decaying particle.
           } 
 \label{fig:ang_def} 
 \end{figure}

Angular variables which can be used
in the analysis of CP-properties are defined in Fig.~\ref{fig:ang_def}.
%
%
%
%
To test CP-properties of the Higgs-bosons one can use the 
distributions of the polar  angles $\Theta_1$ and $\Theta_2$
as well as the $\Delta \phi$ distribution, where $\Delta \phi$ 
is the angle between two $Z$- or two $W$- decay planes.
To simplify the analysis, 
instead of two-dimensional distribution in ($\cos \Theta_1$, $\cos \Theta_2$)
we consider the  distribution in a new variable, defined as
\begin{eqnarray}
 \zeta & = & \frac{\sin^2\Theta_1\;\cdot\; \sin^2\Theta_2}
                 {(1+\cos^2\Theta_1)\cdot (1+\cos^2\Theta_2)} \; .
\end{eqnarray}
The $\zeta$-distribution  corresponds to the ratio of the angular distributions expected for
the decay of a scalar and a pseudoscalar  
(in a limit $M_h >\!\!> M_Z$) \cite{miller}.

\begin{figure}[p]
  \begin{center}
     \epsfig{figure=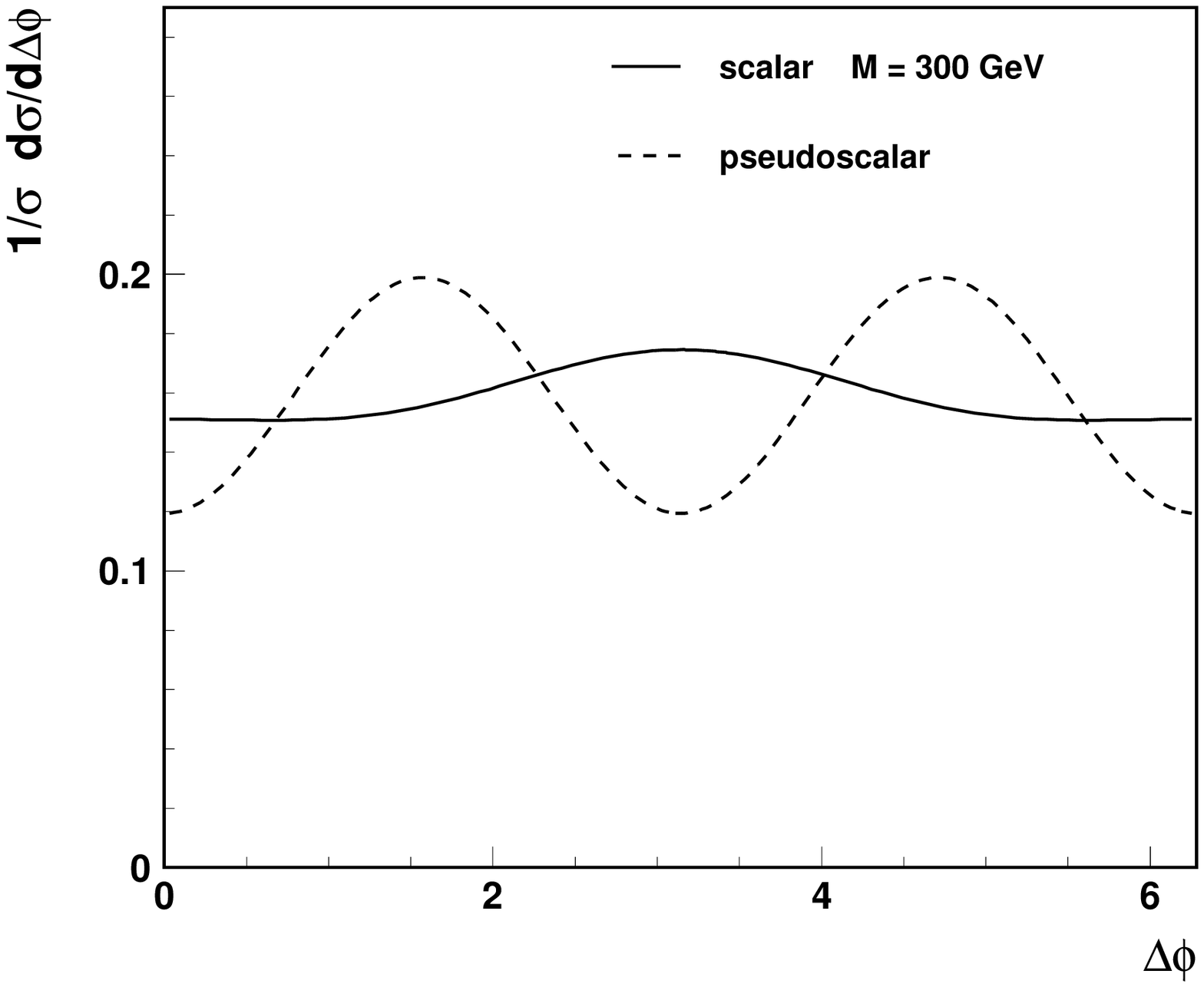,height=\twofigheight,clip=}
     \epsfig{figure=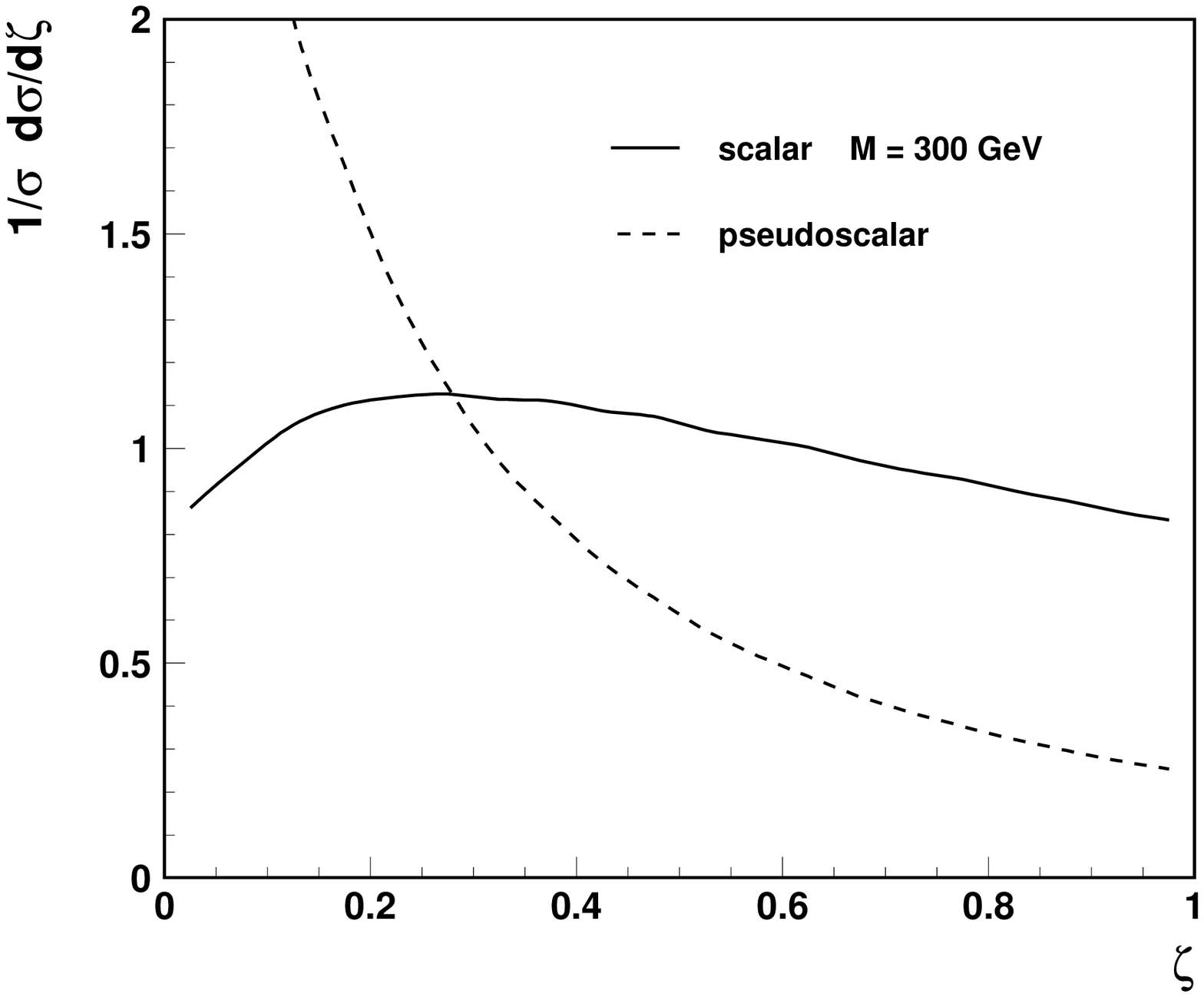,height=\twofigheight,clip=}
  \end{center}
 \caption{
Normalized angular distributions in $\Delta \phi$ (upper plot) and 
$\zeta$ (lower plot), expected (Eq.~\ref{ha}) 
for scalar and pseudoscalar higgs decays 
$H,A \rightarrow Z Z \rightarrow l^+ l^- j j $,
for the  higgs mass of 300 GeV.
           } 
\label{fig:ang_exp} 
 \end{figure}

The angular distributions in $\Delta \phi$ and $\zeta$, expected for
a scalar and a pseudoscalar higgs (Eq.~\ref{ha})  
\mbox{$h \rightarrow Z Z \rightarrow l^+ l^- j j $} are compared
in Fig.~\ref{fig:ang_exp}.
Both distributions clearly distinguish between decays of scalar and
pseudoscalar higgs.
For the scalar  higgs the distributions are almost flat in both
$\Delta \phi$ and $\zeta$, whereas the pseudoscalar coupling 
(see Eq.~\ref{ha}) introduces
a significant $\Delta \phi$ and $\zeta$ dependence.
From the measurement of these distributions we can try to establish the
CP properties of the Higgs boson, 
even without taking into account the production mechanism.
Such measurement should  be considered as the most general one, 
as in principle some unknown heavy charged particles can significantly 
modify the production vertex, affecting the invariant-mass distributions
on which the measurements of the 
two-photon width and two-photon phase are based.

Note, that the distribution of the higgs decay-angle $\Theta_h$ is expected 
to be flat both for scalar and pseudoscalar higgses.
However, the corresponding angular distribution 
it is not flat for the non-resonant background. Therefore the observed 
$\Theta_h$ distributions, where obviously a signal, background and possible 
interference between them contribute, maybe be sensitive to the 
Higgs-boson couplings. 
%


\subsection{Generic model with CP violating couplings}

Following the analysis described in \cite{miller}
we consider a generic higgs model with the following tensor couplings of a 
Higgs boson, denoted below by $\cal H$,  to $ZZ$ and $W^+W^-$:
\begin{eqnarray}
g_{{\cal H}ZZ} & = & i g \frac{M_Z}{\cos \theta_W} \; \left(
  \lambda_H \cdot g^{\mu \nu} \; + \; 
  \lambda_A \cdot \varepsilon^{\mu \nu \rho \sigma}\; 
  \frac{(p_1+p_2)_\rho \; (p_1-p_2)_\sigma}{M_Z^2} \right)\\
g_{{\cal H}WW} & = & i g M_W \; \left(
  \lambda_H \cdot g^{\mu \nu} \; + \; 
  \lambda_A \cdot \varepsilon^{\mu \nu \rho \sigma}\; 
  \frac{(p_1+p_2)_\rho \; (p_1-p_2)_\sigma}{M_W^2} \right)
\end{eqnarray}
The Standard Model couplings are reproduced for $\lambda_H = 1$ 
and $\lambda_A = 0$.
Below we will consider only small deviations from the Standard Model, i.e. 
 $\lambda_H \sim 1$ and $|\lambda_A| \ll 1$.
To simplify the case, we assume that higgs ${\cal H}$ couplings to
fermions are the same as in the Standard Model.

\subsection{Reconstruction of the angular distributions}

\begin{figure}[p]
\begin{center}
\epsfig{figure=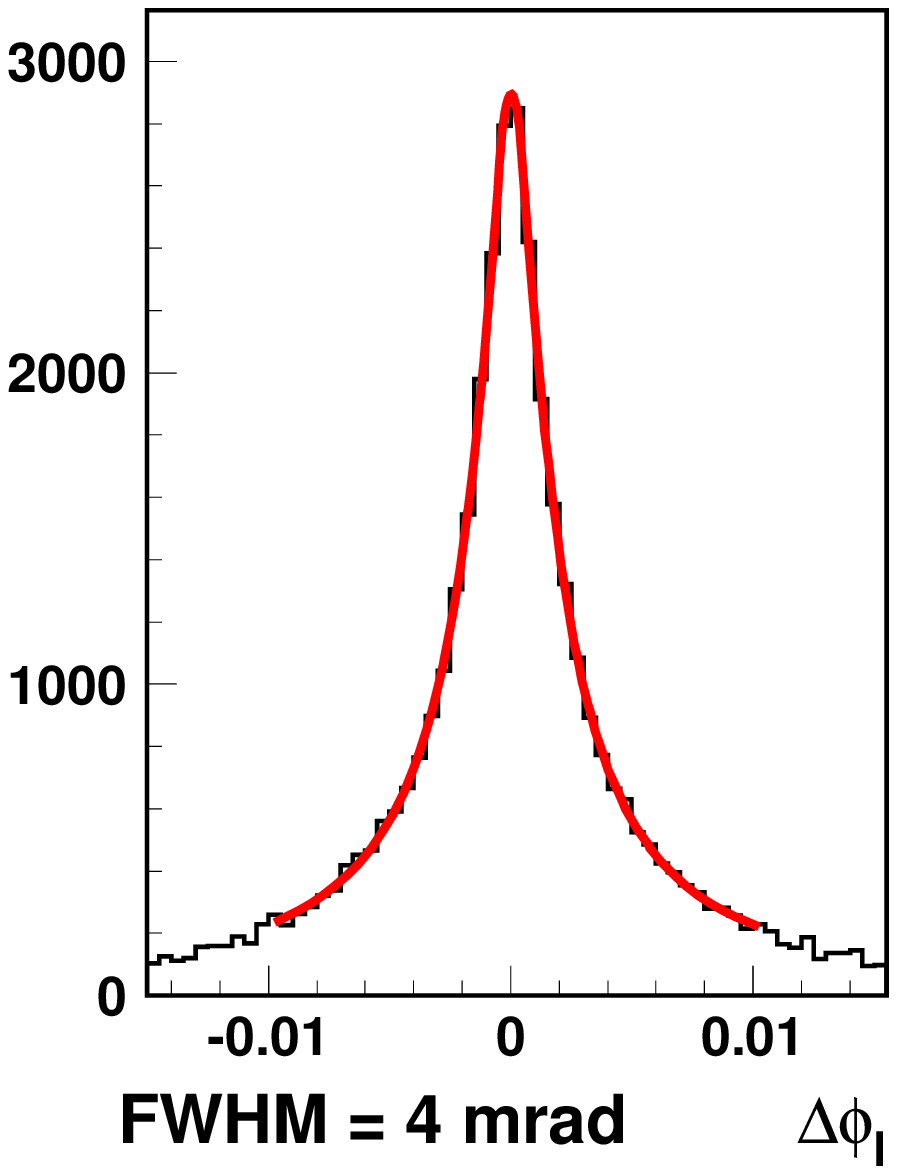,width=4cm,clip=}
\epsfig{figure=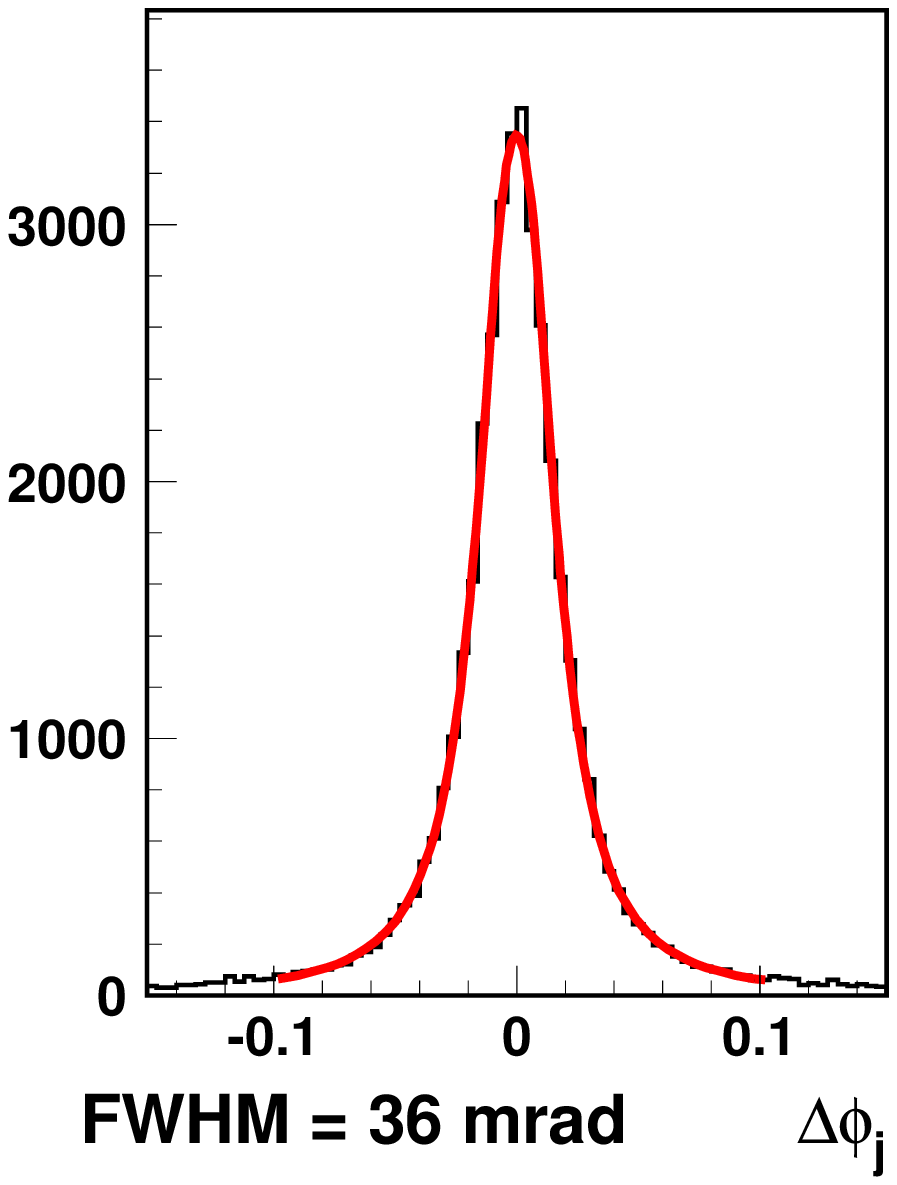,width=4cm,clip=}
\epsfig{figure=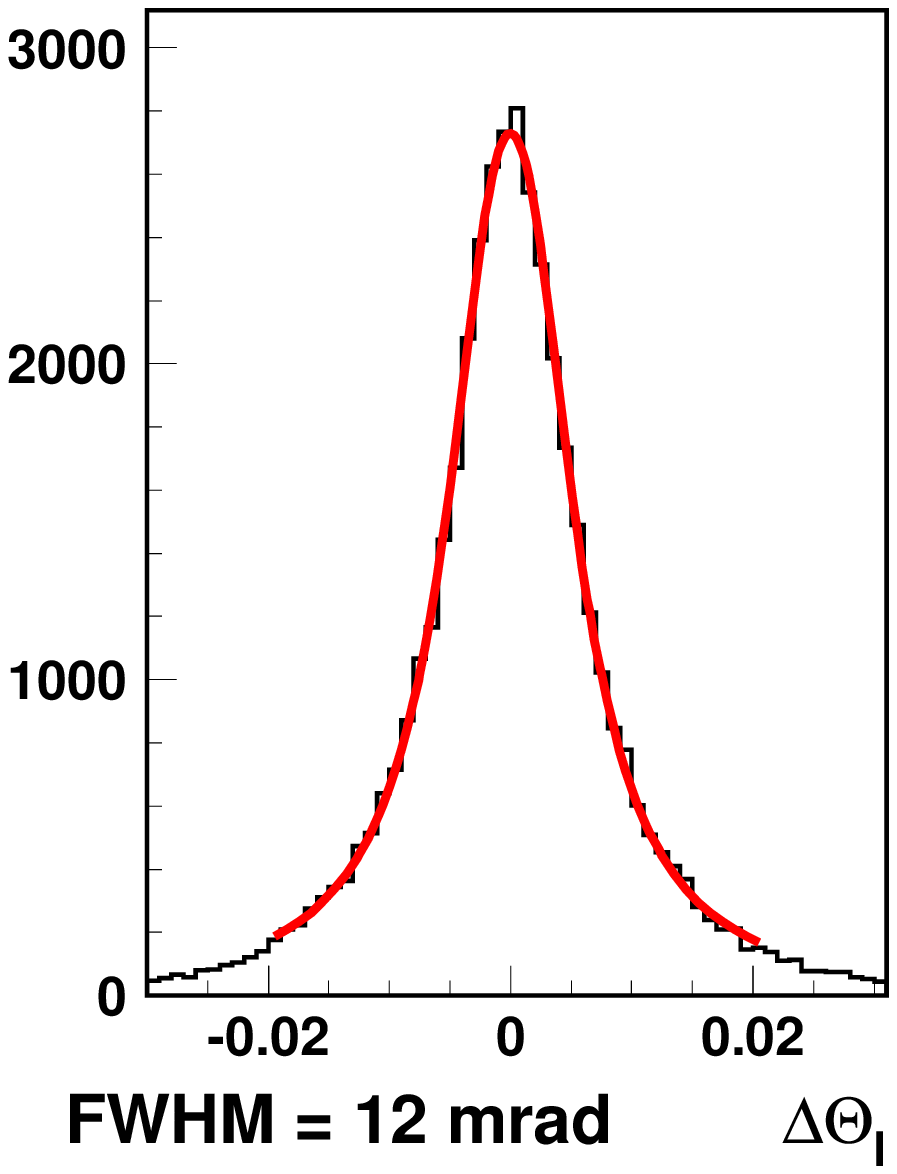,width=4cm,clip=}
\epsfig{figure=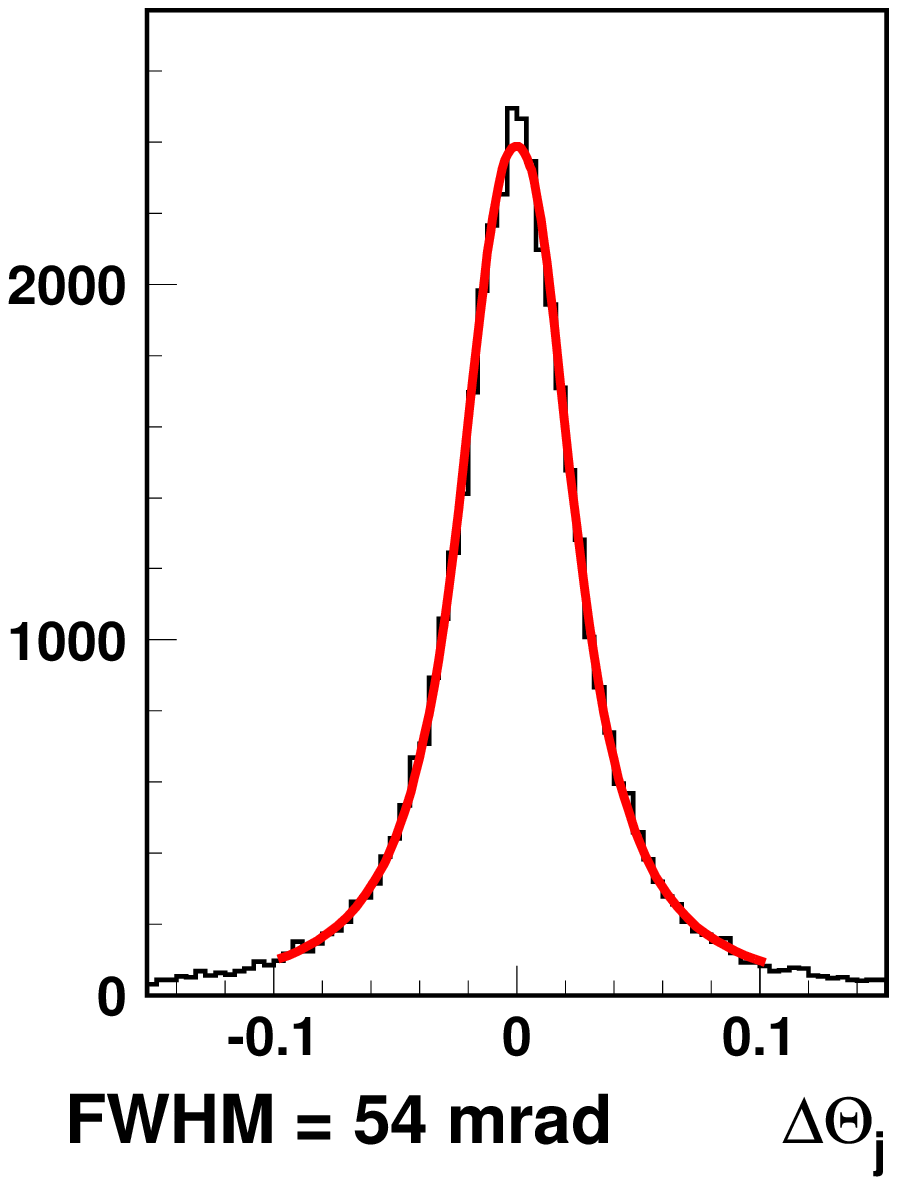,width=4cm,clip=}
\end{center}
 \caption{
     Resolution in the reconstructed $Z$-decay angles $\Theta$ and $\phi$, 
  for the leptonic ($\Theta_l$, $\phi_l$)  and hadronic ($\Theta_j$, $\phi_j$)
     final states.
           Events were simulated with the PYTHIA and SIMDET programs, for
           a primary electron-beam energy of 209 GeV and scalar
          Higgs-boson mass of 300~GeV.
           } 
 \label{fig:ang_res} 
 \end{figure}

Measurement of the angular distributions has been studied 
using samples of $ZZ$ events generated with PYTHIA and SIMDET 
programs, as described in section \ref{sec:width}.
The resolutions of the reconstructed decay angles $\Theta$ and $\phi$, 
for leptonic  and hadronic $Z$ decays, are compared in
Fig.~\ref{fig:ang_res}, for the scalar Higgs-boson with mass of 300~GeV.
%

%
%
%

\begin{figure}[p]
  \begin{center}
     \epsfig{figure=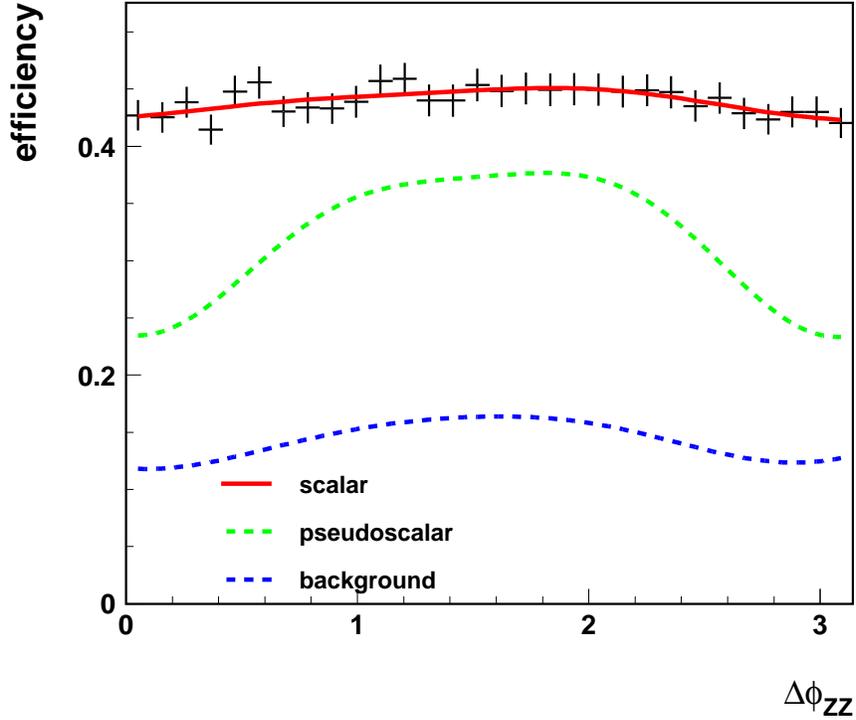,height=\figheight,clip=}
  \end{center}
 \caption{ 
         Selection efficiency as a function of the angle $\Delta \phi$
        between two $Z$ decay planes, for 
        $Z Z \rightarrow l^+ l^- j j $ events
         coming from the scalar higgs decays, pseudoscalar higgs decays 
         and non-resonant background.
         Events were simulated with the PYTHIA and SIMDET programs, for
         a primary electron-beam energy of 209 GeV and 
         Higgs-boson mass of 300~GeV.
      Only events with the reconstructed higgs mass between 286 and  312 GeV
         are accepted.
         } 
 \label{fig:acc_zz_300} 
 \end{figure}

A very good resolution is obtained for all considered distributions,
nevertheless
the measured angular distributions are strongly affected by 
the selection cuts used in the analysis.
Because of the cut on the lepton and jet angles, applied
to preserve a good mass resolution (see \cite{nzk_wwzz} for more details),
we observe significant loss of the selection efficiency for
events with lepton or jet emitted along the beam direction.
Also the Durham jet algorithm used in the event reconstruction
imposes constraints on the angular separation between leptons and jets.
Both effects result in a highly non-uniform angular acceptance.
Selection efficiencies 
for $\gamma \gamma \rightarrow Z Z \rightarrow l^+ l^- j j $ events,
as a function of the angle $\Delta \phi$,
are presented in Fig.~\ref{fig:acc_zz_300}.
We consider reconstructed $\Delta \phi$ values  from 0 to $\pi$, since 
we are not able to distinguish between quark and antiquark jet.
The efficiencies for $ZZ$ events coming from  decays of scalar, 
pseudoscalar higgs   and non-resonant background
are compared.
The polar-angle distributions differ among these three classes of events, what leads to a significantly different acceptances.
This simulation was performed for primary electron-beam energy of 209~GeV 
optimal for a Higgs-boson mass of 300~GeV.
An additional cut on the reconstructed $ZZ$ invariant-mass has been
introduced to optimise the signal measurement from  the angular distribution.
We found that for higgs mass of 300~GeV the accepted mass range lies 
between 286 and  312~GeV.
This cuts affects mainly the non-resonant background events.

%
%

We calculate the expected angular distributions for $ZZ$ and
$W^+ W^-$ events by convoluting the corresponding cross-section formula with 
the CompAZ photon-energy spectra, and in addition with the 
parametrization of the invariant-mass 
resolution and acceptance function including the angular- and jet-selection 
cuts.
Here we neglect (small) effects of the angular resolution.
For the measurement of the event distributions in 
$\Delta \phi$ and $\zeta$ we introduce the  additional cuts 
on the reconstructed $ZZ$ invariant mass (for $ZZ$ events)
or on the reconstructed  $W^+ W^-$ invariant mass as well as the
higgs-decay angle $\Theta_h$ (for $W^+ W^-$ events).
The cuts were optimised for the smallest relative error in 
the signal cross-section measurement.

\begin{figure}[p]
  \begin{center}
     \epsfig{figure=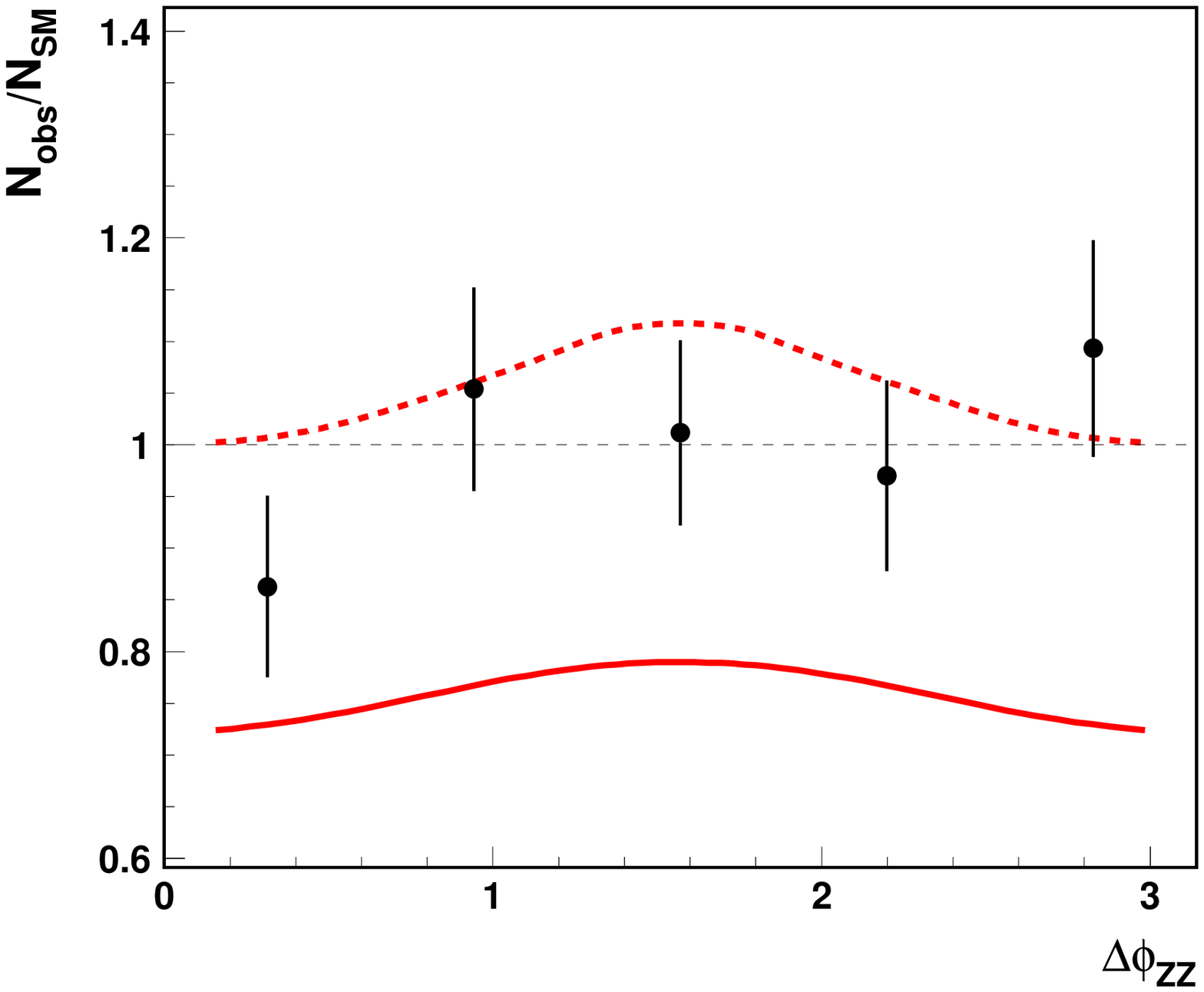,height=\twofigheight,clip=}
     \epsfig{figure=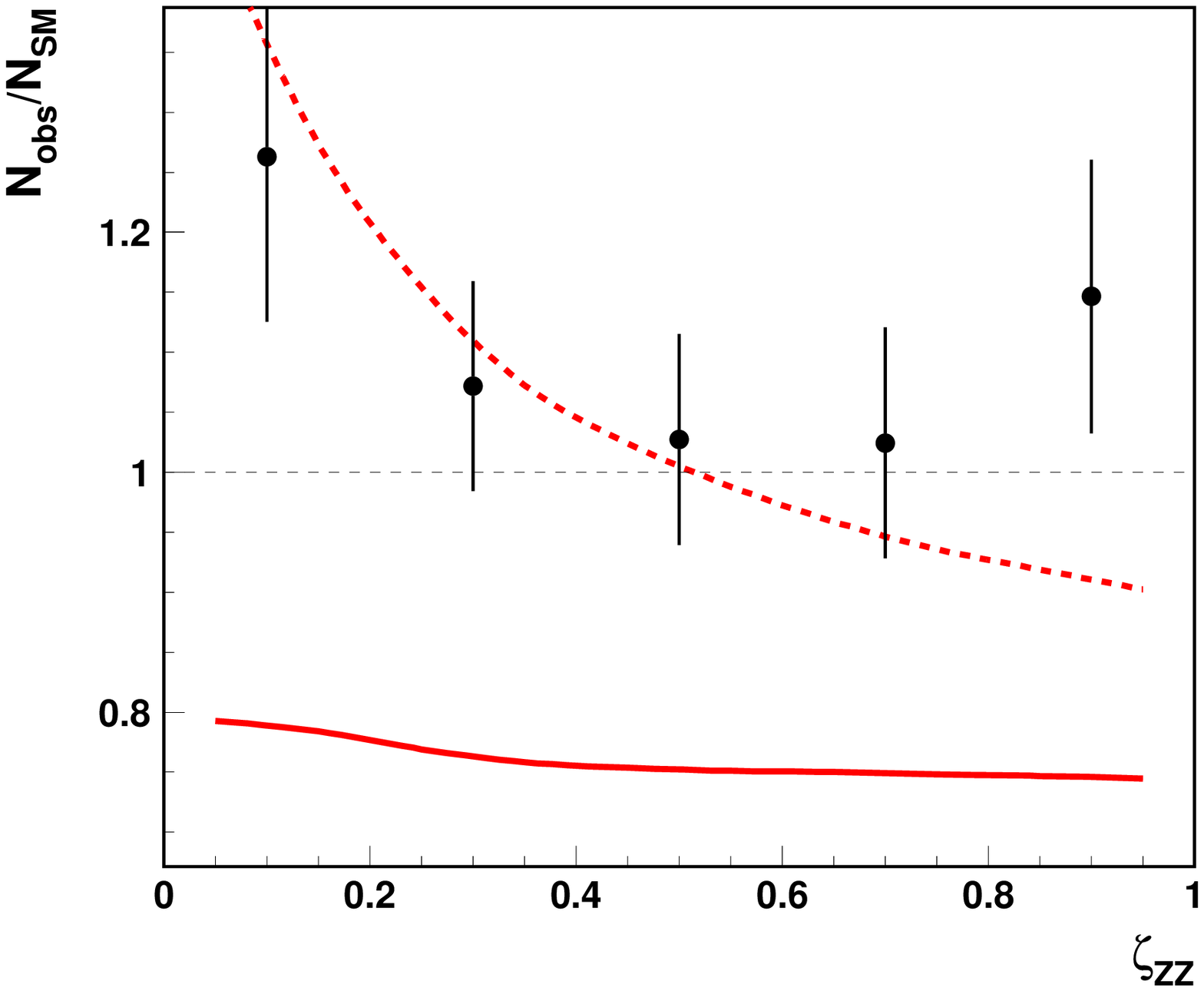,height=\twofigheight,clip=}
  \end{center}
 \caption{ 
         Expected deviations from the Standard Model predictions
        for  $Z Z \rightarrow l^+ l^- j j $ events,
        for the measurement of the angle $\Delta \phi_{ZZ}$
        between two $Z$-decay planes (upper plot) and
        of the variable $\zeta_{ZZ}$ calculated
        from the polar angles of the $Z\rightarrow l^+ l^-$
        and  $Z\rightarrow j j $  decays (lower plot).
         Solid (dashed) red line correspond to the model with $\lambda_H = 1$
         and  $\lambda_A = 0.2$ ($\lambda_A = -0.2$).
         Signal and background calculations are performed for
         primary electron-beam energy of 209 GeV and 
         the Higgs-boson mass of 300~GeV.
         Only events with reconstructed higgs mass between 286 and  312 GeV
         are accepted.
         } 
 \label{fig:histzz} 
 \end{figure}

\begin{figure}[p]
  \begin{center}
     \epsfig{figure=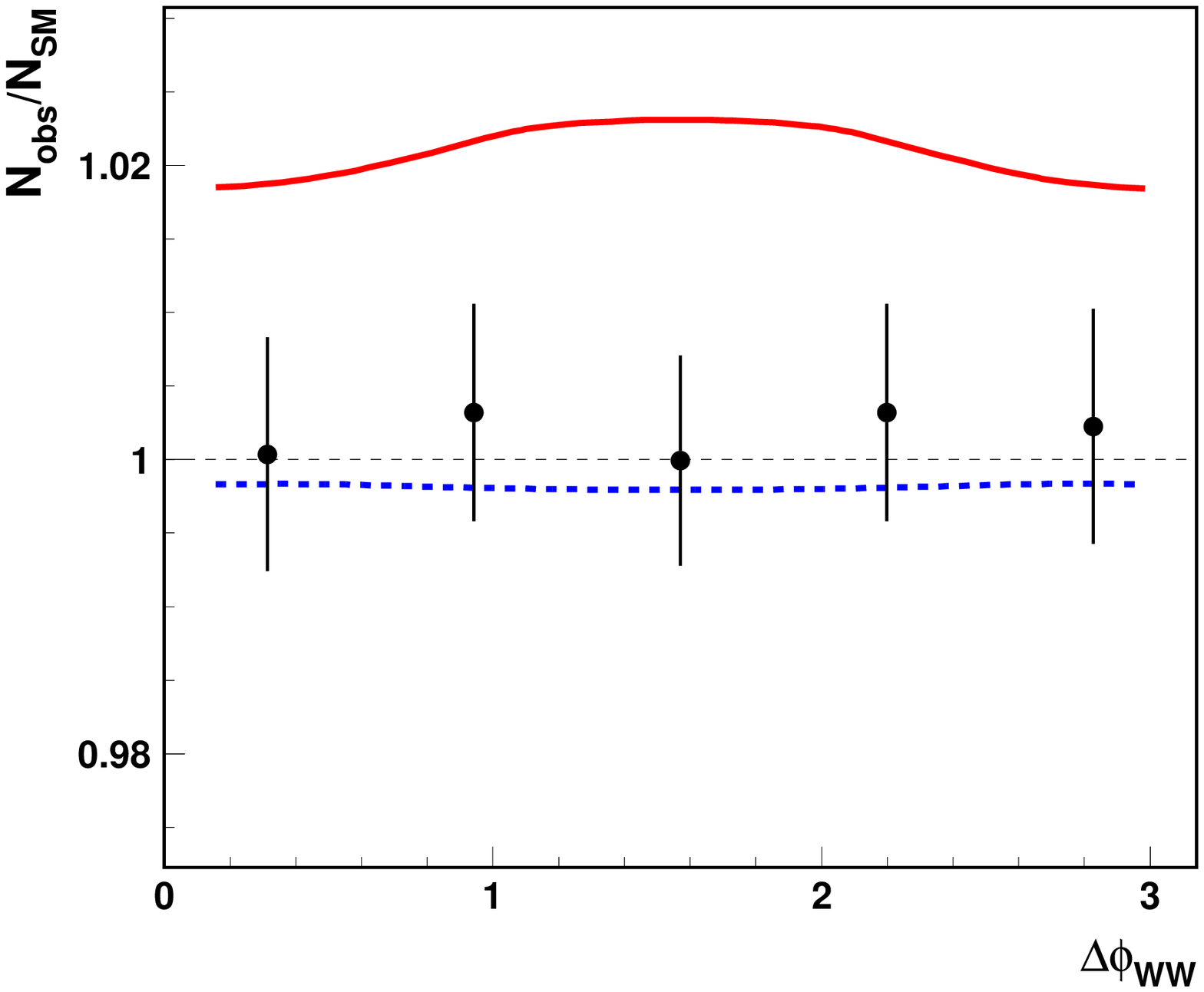,height=\twofigheight,clip=}
     \epsfig{figure=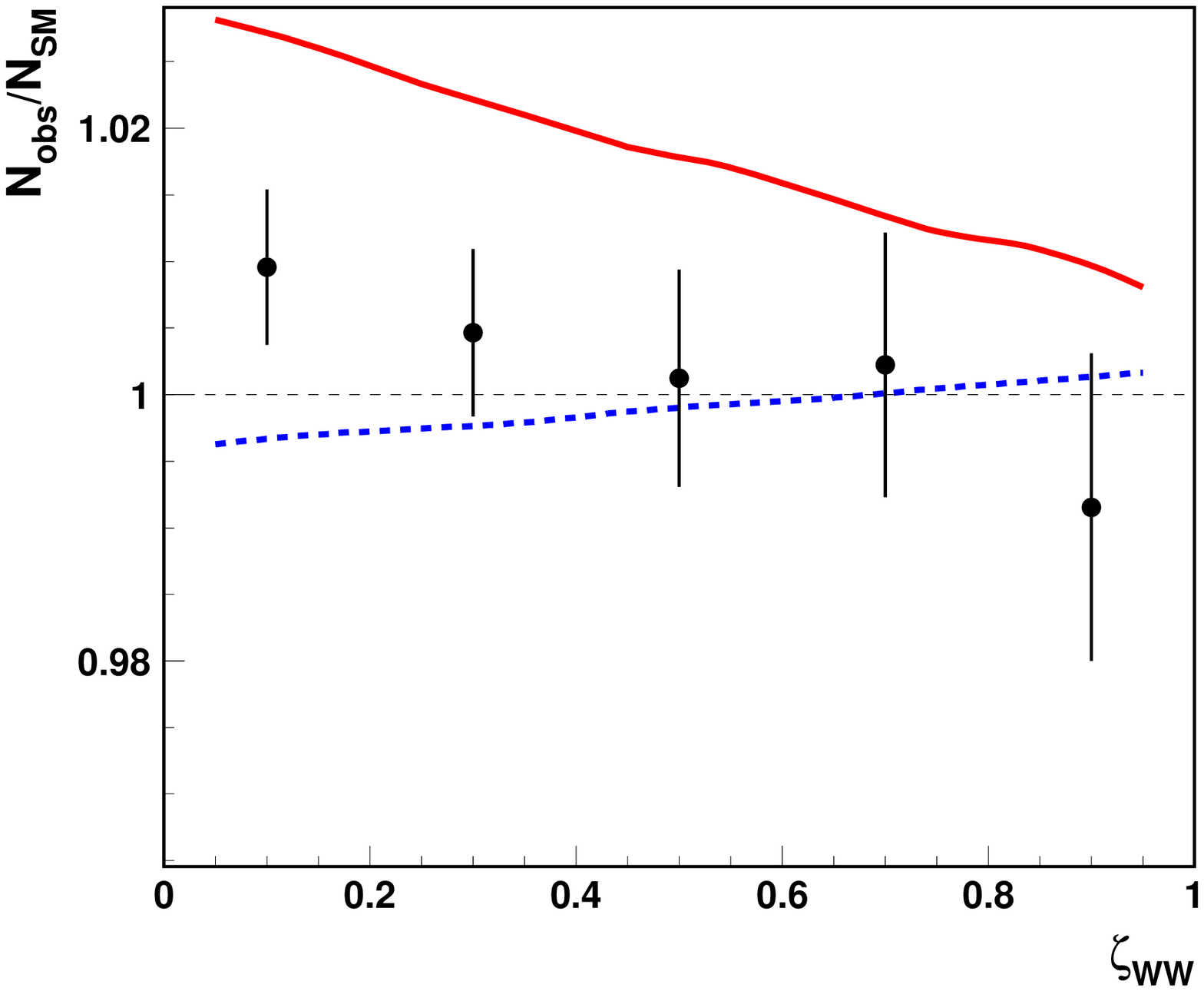,height=\twofigheight,clip=}
  \end{center}
 \caption{ 
         Expected deviations from the Standard Model predictions
         for  $W^+ W^- \rightarrow 4 j $ events,
        for the measurement of the angle $\Delta \phi_{WW}$
        between two $W$-decay planes (upper plot) 
        and of the variable  $\zeta_{WW}$ calculated
        from the polar angles of two $W\rightarrow  j j $  decays
        (lower plot).
         Solid red (dashed blue) line correspond to the model with
          $\lambda_A = 0.2$ and $\lambda_H = 1$ 
          ($\lambda_A = 0$ and $\lambda_H = 1.1$).
         Signal and background calculations are performed for
         primary electron-beam energy of 209 GeV and 
         the Higgs-boson mass of 300~GeV.
       Correlated cuts on the reconstructed  mass and higgs decay angle were
         imposed to improve signal to background ratio.
         } 
 \label{fig:histww} 
 \end{figure}

Expected precision in the measurements of the angle $\Delta \phi$  
and of the variable $\zeta$  distributions, for 
$\gamma \gamma \rightarrow Z Z \rightarrow l^+ l^- j j $ events
and $\gamma \gamma \rightarrow W^+ W^- \rightarrow 4  j $ events
are illustrated in Figs.~\ref{fig:histzz}  and \ref{fig:histww}, respectively. 
%
%
Calculations were performed for
primary electron-beam energy of 209 GeV and 
the Higgs-boson mass of 300~GeV.
%
%
The results, presented relative to the SM prediction, 
are confronted in Figs. \ref{fig:histzz} (\ref{fig:histww}) 
with the expectations of the generic model with
$\lambda_A = \pm 0.2$ and $\lambda_H = 1$
($\lambda_A = 0.2$ and  $\lambda_H = 1$;
 $\lambda_A = 0$ and $\lambda_H = 1.1$ ), 
for the reconstructed  $ZZ$ ($W^+ W^-$) events.
We see, that the generic model predicts deviations both in the normalization
and in the shape of angular distributions.
Therefor we should be able to constrain Higgs-boson couplings from
the shape of the distributions, event if the overall normalization
related to the Higgs-boson production mechanizm is not known.

%
%
%
%
%

\subsection{Determination of CP properties of the Higgs boson \\
           from the measurement of angular distributions}

\begin{figure}[p]
  \begin{center}
     \epsfig{figure=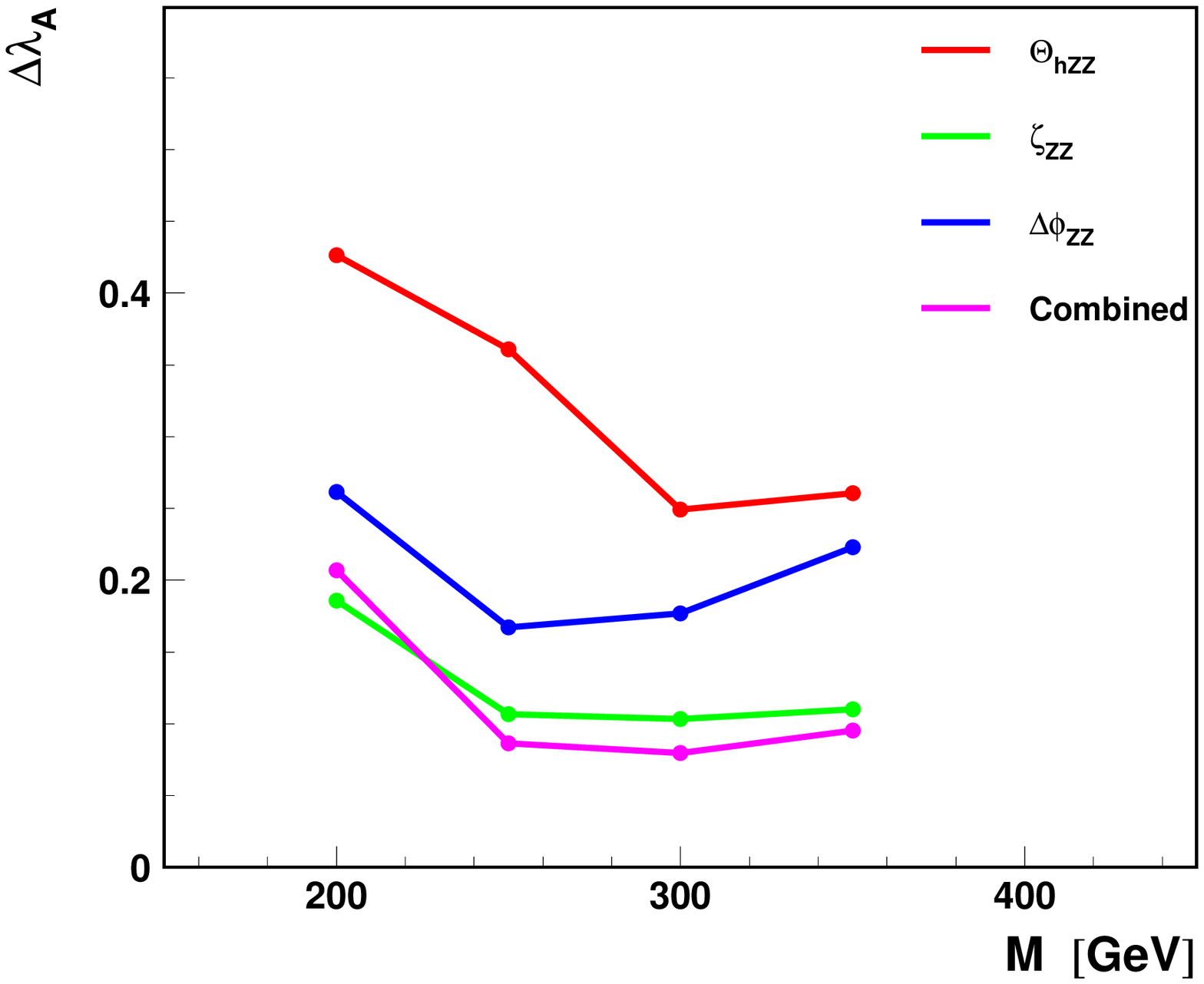,height=\twofigheight,clip=}
     \epsfig{figure=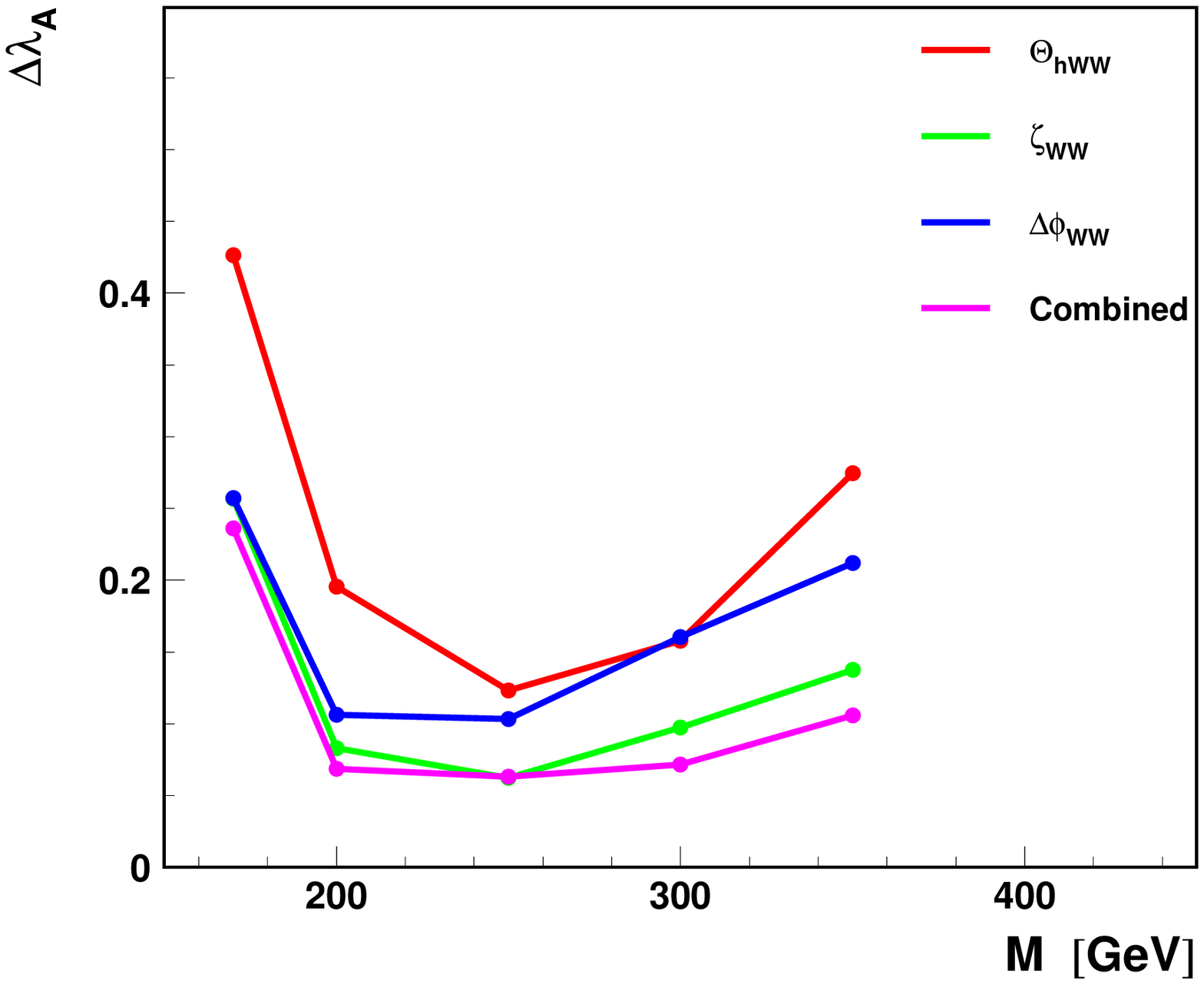,height=\twofigheight,clip=}
  \end{center}
 \caption{ 
          Statistical error in the determination of 
$\lambda_A$,  expected after one year of photon collider running,  
as a function  of the Higgs-boson mass M.
Fits were performed to the shape of three individual angular distributions,
as indicated in the plots, separately
for $ZZ$ events (upper plot) and $W^+ W^-$  events (lower plot).
Results of the simultaneous fits to all three considered distributions 
are also given.
Parameter $\lambda_H$ is fixed to the Standard Model value, $\lambda_H = 1$.
No normalisation constraints are imposed.
Errors resulting from the fit were calculated for $\lambda_A = 0$.
%
         } 
 \label{fig:resazzww} 
 \end{figure}

Each of the considered angular distributions discussed above
 can be fitted with a model expectations, given in terms of the parameters 
$\lambda_A$ and $\lambda_H$.
If we do not use normalisation constraints (i.e. do not assume
a given production mechanism) there are large correlations
between $\lambda_A$ and $\lambda_H$ in fits to a single-variable 
 distributions, i.e.  for  $\Delta \phi$, $\zeta$ and $\Theta_h$.
Therefore we first limit ourselves to $\lambda_H = 1$ case.
Statistical errors in the determination of 
$\lambda_A$ from the fits to the shape of various angular distributions,
to be obtained  after one year of photon collider running,  
are presented in Fig.~\ref{fig:resazzww}, separately for the $WW$ and $ZZ$ 
channels.
In the figure we present also the results obtained from
 the simultaneous (combined) fit to all three considered distributions.
It turns out that both for $ZZ$ and for $W^+ W^-$  events 
the measurement of the $\zeta$-distribution gives the best 
constraints on the $\lambda_A$ parameter, and
a precision of $\lambda_A$ measurement from the $\zeta$ distribution only
is close to the precision obtained from the combined fit.
For $M_h \ge $250~GeV the expected accuracy for $\lambda_A$  is about 0.1 
both for the $ZZ$ and for $W^+ W^-$  events.

%
%

\begin{figure}[tb]
  \begin{center}
     \epsfig{figure=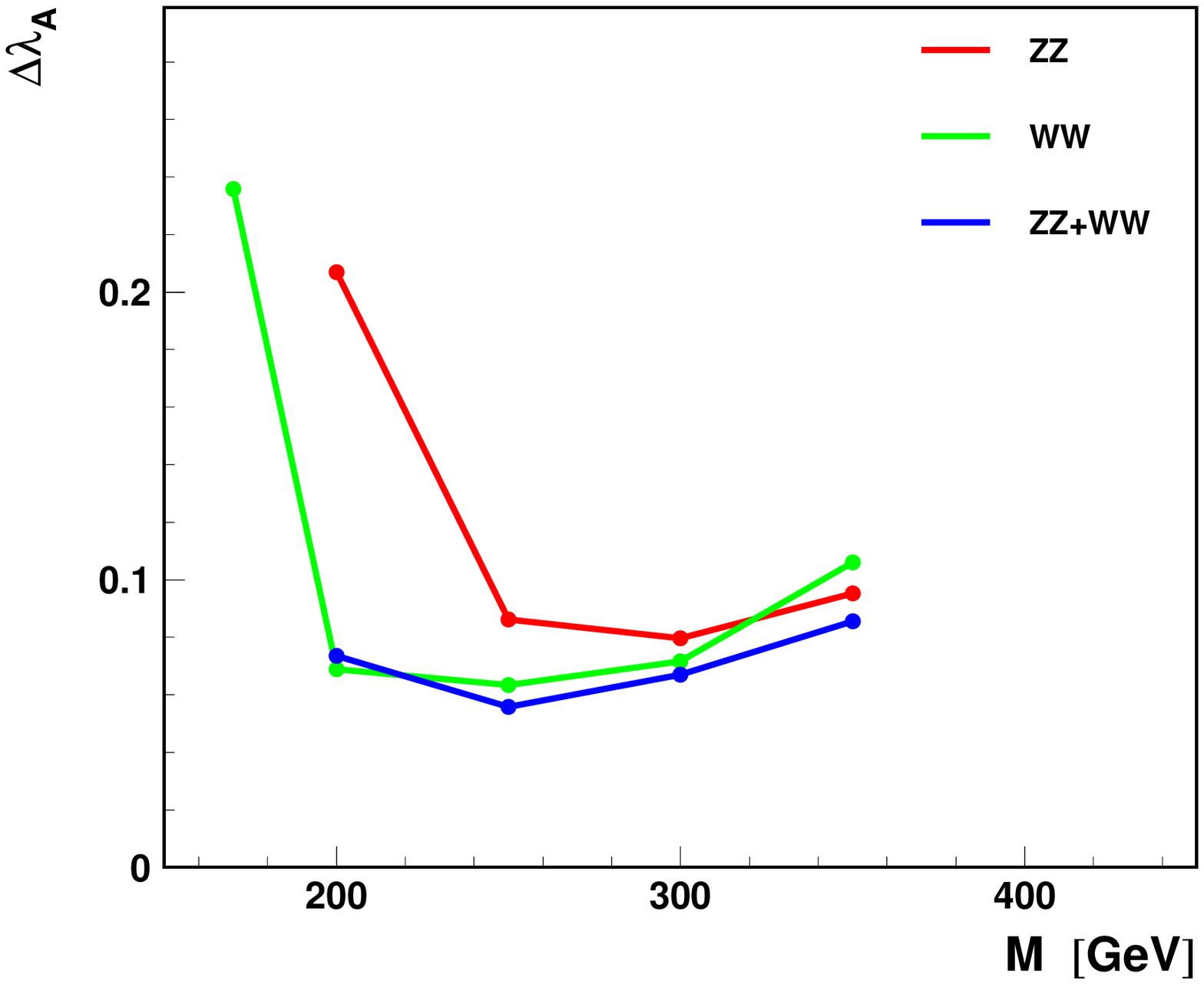,height=\figheight,clip=}
  \end{center}
 \caption{ 
          Statistical error in the determination of 
$\lambda_A$,  expected after one year of photon collider running,  
as a function  of the Higgs-boson mass M.
Results from the simultaneous fits to all considered angular distributions
for $ZZ$ and $W^+ W^-$  events, and from the combined fit 
for both $ZZ$ and $W^+ W^-$  events are compared.
Parameter $\lambda_H$ is fixed to the Standard Model value, $\lambda_H = 1$.
No normalisation constraints are imposed.
Errors resulting from the fit were calculated for $\lambda_A = 0$.
%
         } 
 \label{fig:resaa} 
 \end{figure}

Finally we estimate the expected statistical errors from the combined fit to 
angular distributions
of $ZZ$ and $W^+ W^-$  events, they are shown in  Fig.~\ref{fig:resaa}.
By fitting all angular distributions simultaneously we found the error on
 $\lambda_A $ of 0.06--0.08  for $M_h \ge$ 200~GeV. 

\begin{figure}[p]
  \begin{center}
     \epsfig{figure=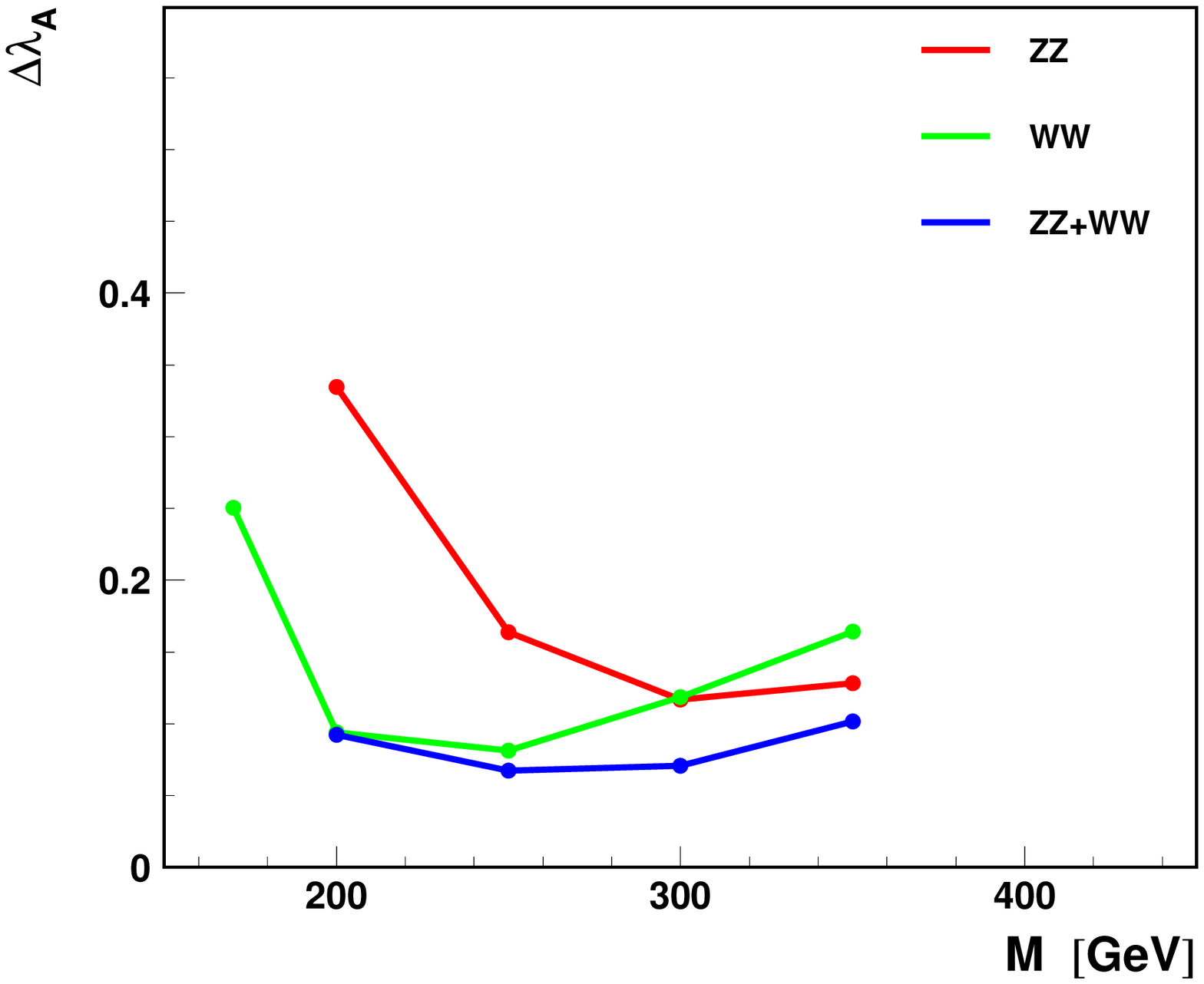,height=\twofigheight,clip=}
     \epsfig{figure=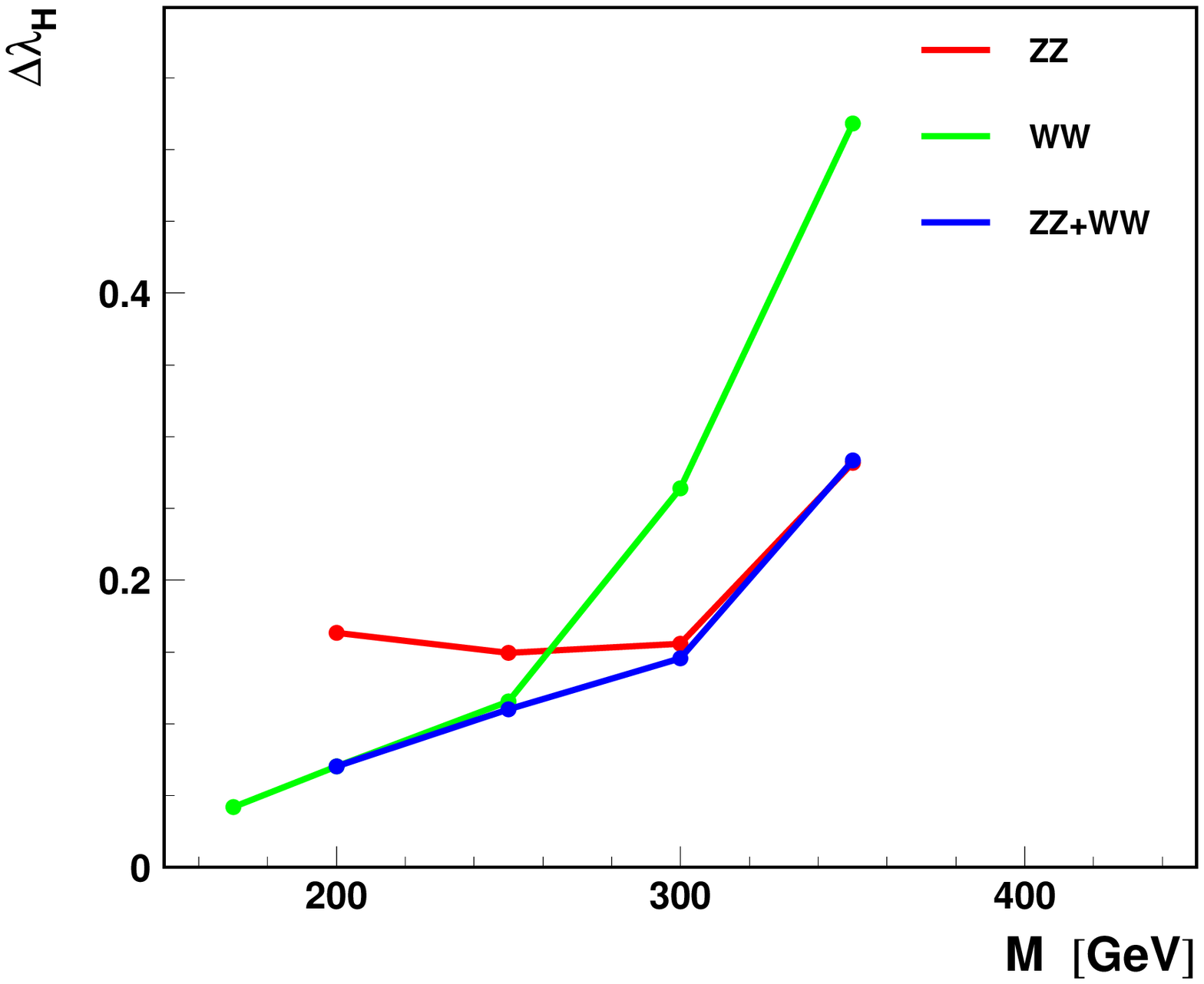,height=\twofigheight,clip=}
  \end{center}
 \caption{ 
          As in \ref{fig:resaa}, for 
$\lambda_A$ (upper plot) and $\lambda_H$ (lower plot),  
for determination of both parameters from simultaneous fit to
all considered angular distributions.
         } 
 \label{fig:resha2} 
 \end{figure}

In case of the combined fit, correlations between  $\lambda_A $ and 
$\lambda_H $ parameters are smaller, so both parameters can be
fitted simultaneously.
Expected statistical errors on $\lambda_A $ and 
$\lambda_H $ are shown in Fig.~\ref{fig:resha2}.
For low Higgs-boson masses, $M_h \le $250~GeV, better constrains are obtained
from the measurement of $W^+ W^-$  events, whereas for $M_h \ge $300~GeV
smaller errors are obtained from the measurement of $ZZ$  events.
Expected error on  $\lambda_A $ is below 0.1 for $M_h \le $200~GeV, while
 the corresponding error on $\lambda_H $ changes from about 0.1 to 0.3 with
increasing Higgs-boson mass.

%
%
%
%

\subsection{Determination of CP properties from the combined\\
       analysis of the invariant-mass and angular distributions}

\begin{figure}[b]
  \begin{center}
     \epsfig{figure=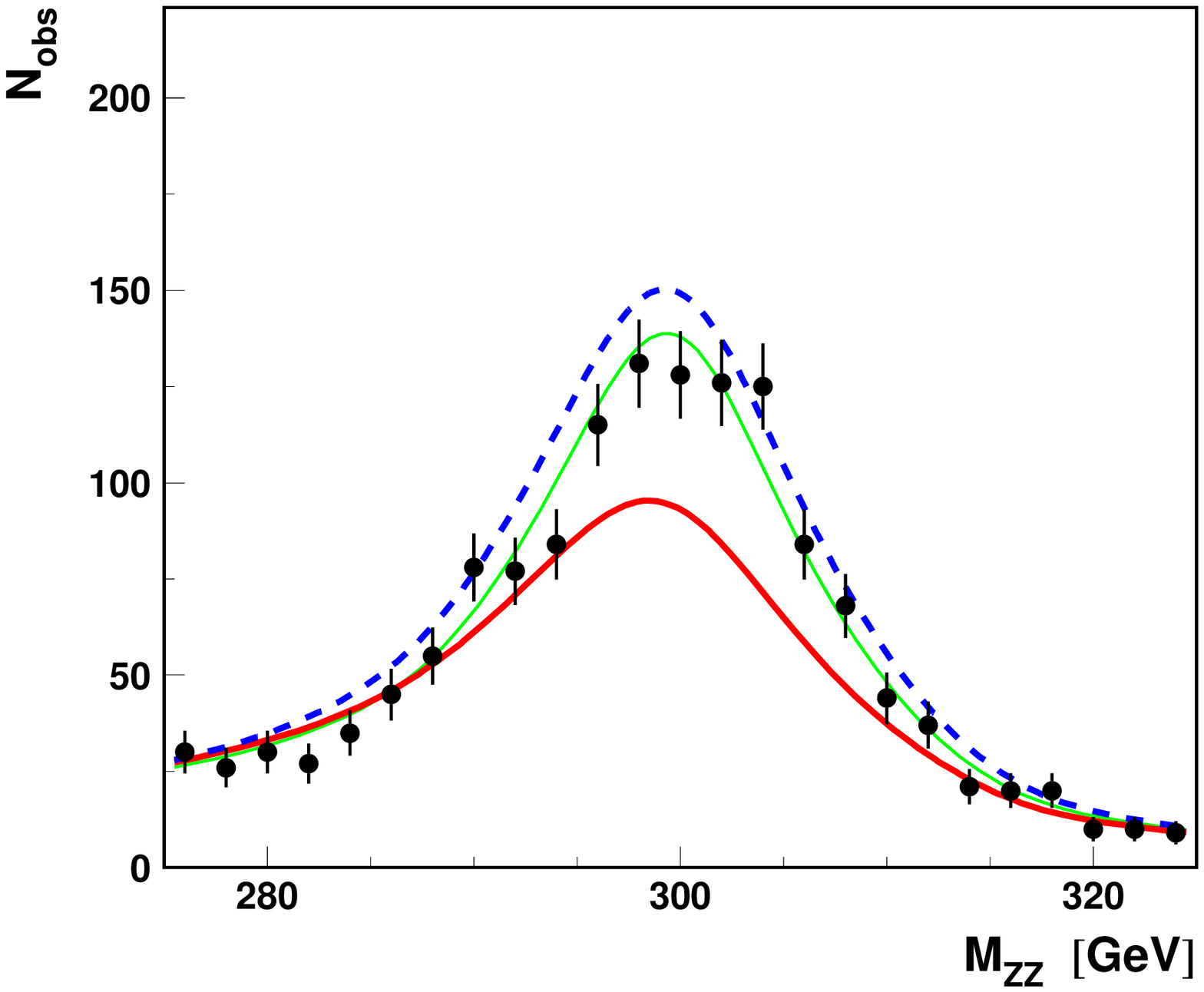,height=\figheight,clip=}
  \end{center}
 \caption{ 
         Predictions
        for the measurement of the $ZZ$ invariant mass $M_{ZZ}$, for
        $Z Z \rightarrow l^+ l^- j j $ events.
         Solid red (dashed blue) line correspond to the generic model with
          $\lambda_A = 0.2$ and $\lambda_H = 1$ 
          ($\lambda_A = 0$ and $\lambda_H = 1.1$).
         Thin green line corresponds to the Standard Model predictions
         ($\lambda_A = 0$ and $\lambda_H = 1$).
         Points with error bars indicate the statistical precision of the
         measurement after one year of photon collider
        running at nominal luminosity. 
         Signal and background calculations are performed for
         primary electron-beam energy of 209 GeV optimal for
         the Higgs-boson mass of 300~GeV.}
 \label{fig:histmzz} 
 \end{figure}

\begin{figure}[p]
  \begin{center}
     \epsfig{figure=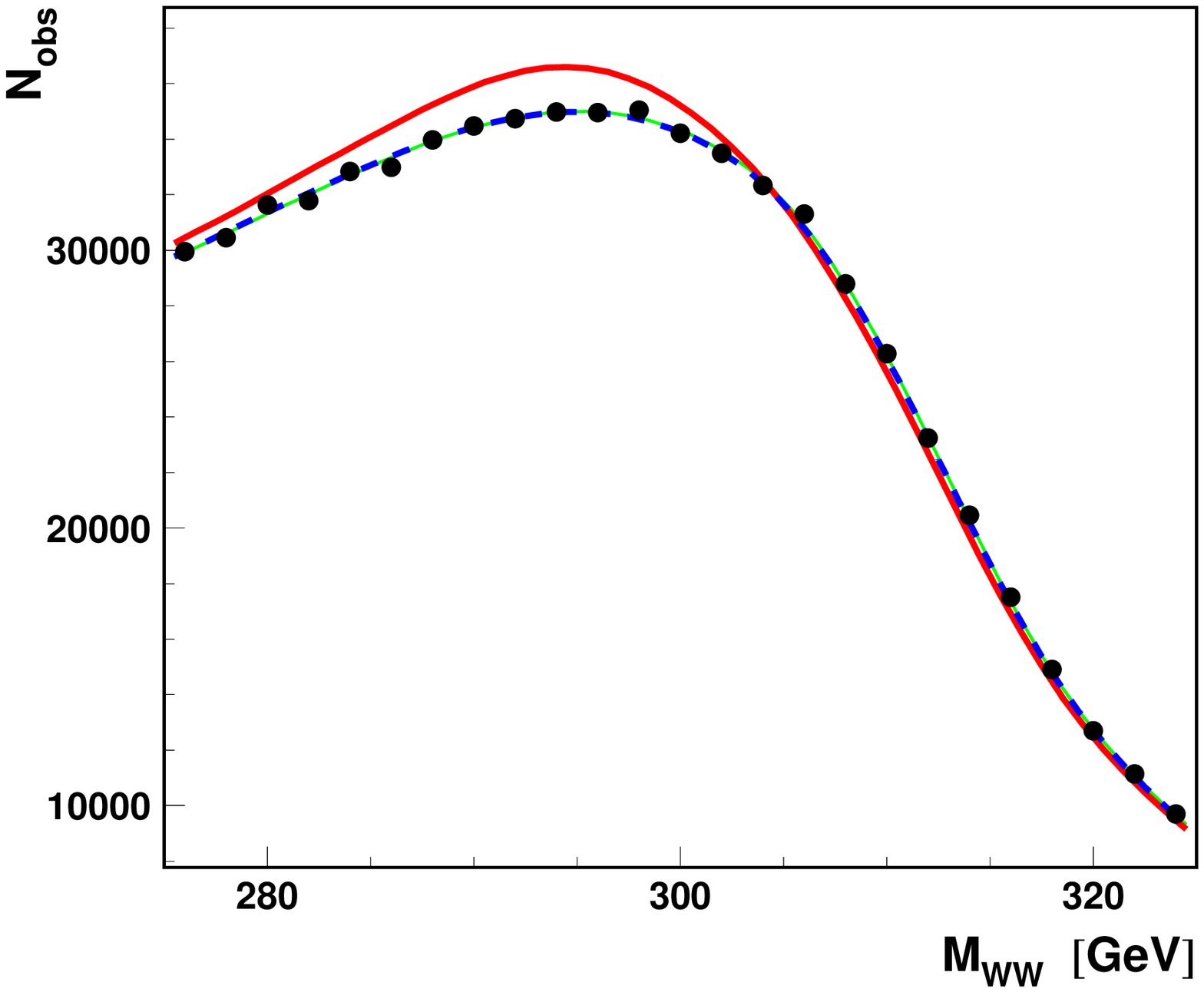,height=\twofigheight,clip=}
     \epsfig{figure=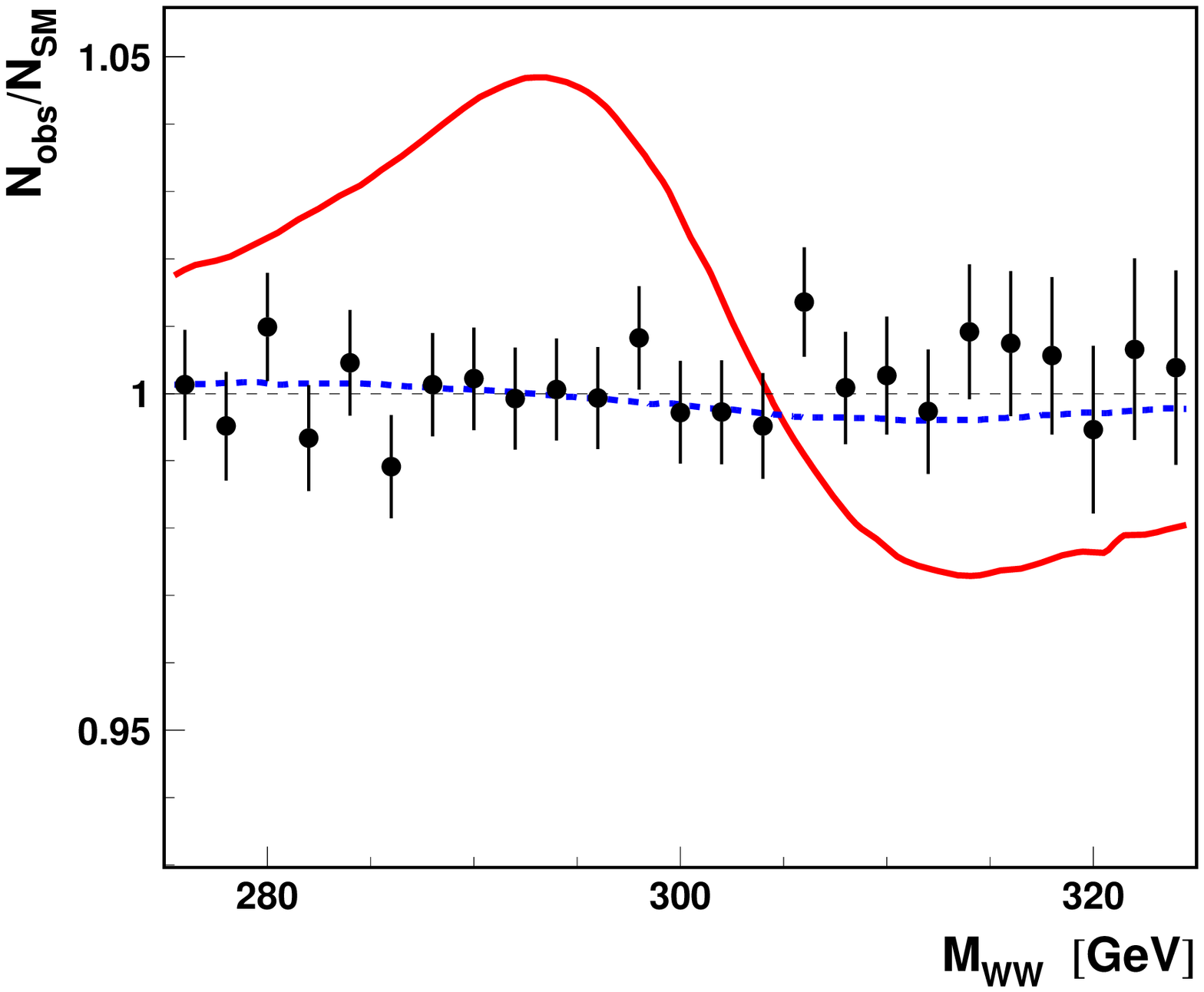,height=\twofigheight,clip=}
  \end{center}
 \caption{ 
         Expected deviations from the Standard Model predictions
        for the measurement of the $W^+ W^-$ invariant mass $M_{WW}$, for
        $W^+ W^- \rightarrow 4 j $ events.
        Invariant-mass distribution measured after one year of 
        photon collider running at nominal luminosity (upper plot)
        and the ratio of this distribution to the SM predictions (lower plot)
        are shown.
         Solid red (dashed blue) line correspond to the model with
          $\lambda_A = 0.2$ and $\lambda_H = 1$ 
          ($\lambda_A = 0$ and $\lambda_H = 1.1$).
         Thin green line (upper plot) corresponds to 
         the Standard Model predictions ($\lambda_A = 0$ and $\lambda_H = 1$).
         Signal and background calculations are performed for
         primary electron-beam energy of 209 GeV optimal
         the Higgs-boson mass of 300~GeV.}
 \label{fig:histmww} 
 \end{figure}

Significantly better constraints on the model parameters $\lambda_A $ 
and $\lambda_H $ can be obtained if  we assume that there 
is no ``new physics'' except for the
considered anomalous couplings of the Higgs-boson to $W^+ W^-$ and $ZZ$.
We can then calculate the expected invariant-mass distributions
for $ZZ$ and $W^+ W^-$ events.
In Fig.~\ref{fig:histmzz} and \ref{fig:histmww}
statistical errors on the measured invariant-mass distributions,
expected after one year of photon collider running at a nominal luminosity,
are compared with the predictions of the generic model.
%
We see, that large deviations from the Standard Model prediction
 are expected for $\lambda_A \ne 0$,  mainly due to the interference effects.

%
%
%
%

\begin{figure}[p]
  \begin{center}
     \epsfig{figure=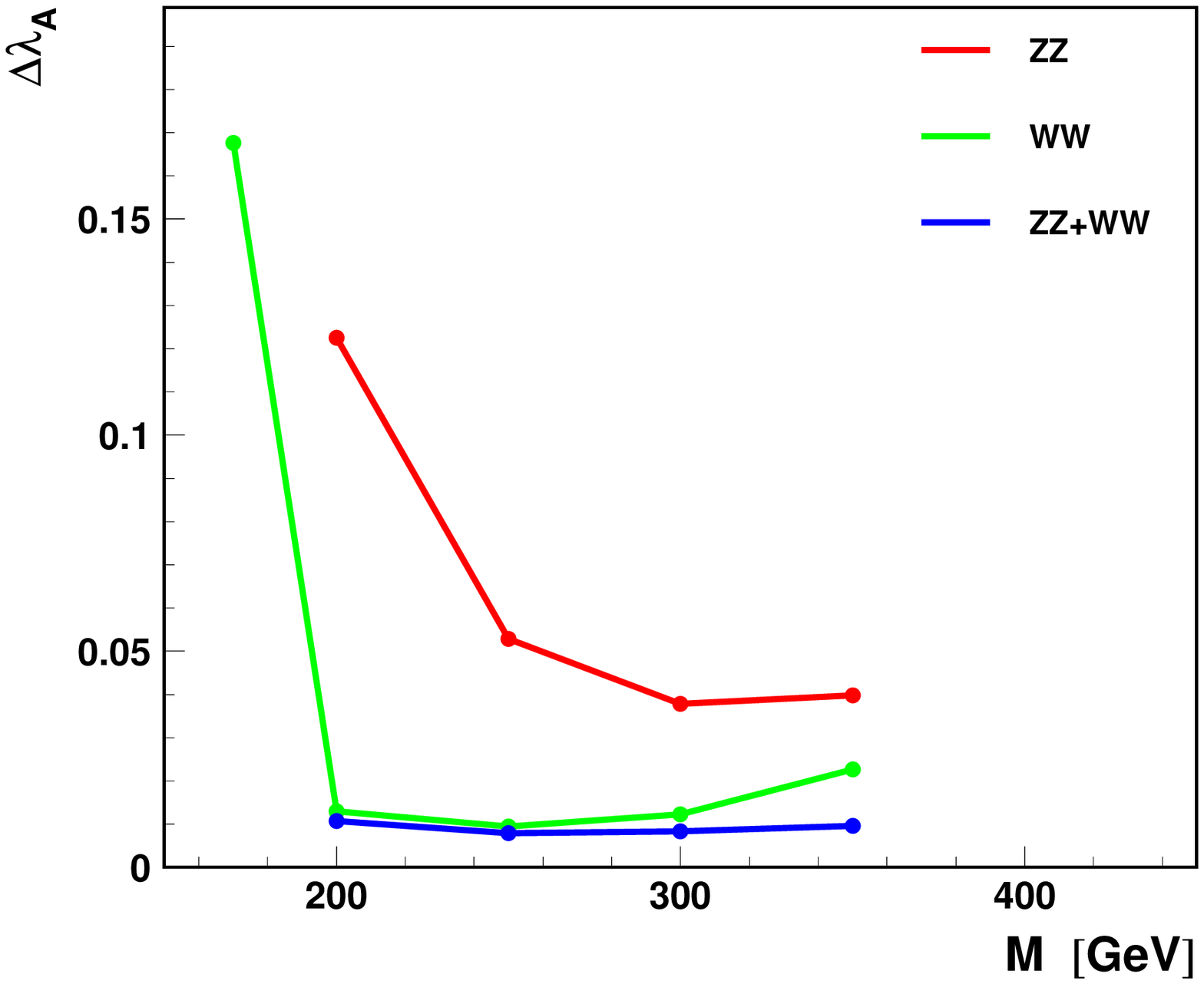,height=\twofigheight,clip=}
     \epsfig{figure=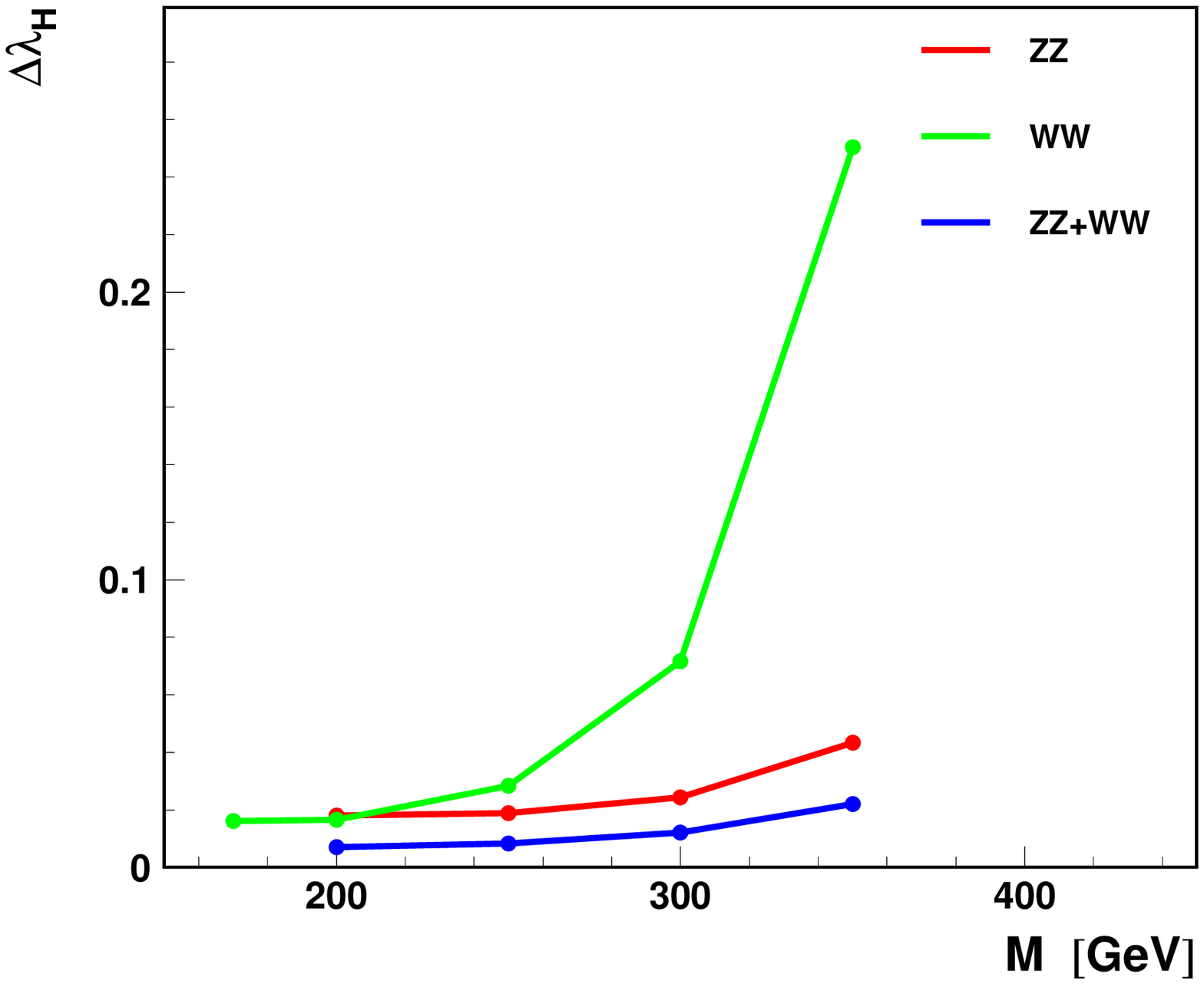,height=\twofigheight,clip=}
  \end{center}
 \caption{ 
          Statistical error in the determination of 
$\lambda_A$ (upper plot) and $\lambda_H$ (lower plot),  
expected after one year of photon collider running,  
as a function  of the Higgs-boson mass M.
Results from the simultaneous fits to all considered angular distributions
and invariant-mass distributions for $ZZ$ and $W^+ W^-$  events, 
and from the combined fit for both $ZZ$ and $W^+ W^-$  events are compared.
Relative normalisation of  $ZZ$ and  $W^+ W^-$ samples
is imposed in the fit.
Errors resulting from the simultaneous fit 
of both parameters  were calculated for $\lambda_A = 0$
and $\lambda_H = 1$.
%
         } 
 \label{fig:resahm} 
 \end{figure}

Expected statistical errors in the determination of $\lambda_A $ and 
$\lambda_H $ from the combined fit to angular distributions
and invariant-mass distributions are compared in Fig.~\ref{fig:resahm}.
A relative normalisation of  $ZZ$ and  $W^+ W^-$ samples is imposed in the fit,
as predicted from the cross section calculations.
However, overall normalisation, corresponding to the integrated luminosity
of the considered data sample, is allowed to vary. 
For $M_h \ge$ 200~GeV the error on $\lambda_A $ of about 0.01
can be achieved from the combined fit to all considered distributions.  
The value of $\lambda_A $ is mainly constrained by the measurement of 
the invariant-mass distribution for $W^+ W^-$  events.
Expected statistical error on $\lambda_H $ is slightly bigger,
between 0.01 and 0.02, and comes mainly from the measurement
of the invariant-mass distribution for $ZZ$  events.

%
%
%

\section{Summary}

An opportunity of measuring of the Higgs-boson properties at the 
the Photon Collider at TESLA has been studied in detail for masses 
between 180 and 350~GeV,
using realistic luminosity spectra and detector simulation. 
We found that from the combined measurement of the invariant-mass 
distributions in the $ZZ$ and $W^+ W^-$ decay-channels, the  parameters 
of the SM-like Two Higgs Doublet Model can be precisely determined.
For both light and heavy scalar Higgs boson, a
statistical precision in the determination of its coupling ($\tan \beta$)
is of the order of 10\% already after one year of Photon Collider running.
In case of the Two Higgs Doublet Model with a CP violation,
the $H-A$ mixing angle can be determined with statistical precision of 
20 to 90~mrad (in a small-mixing approximation).

We also considered the combined measurement of the various
 angular correlations in the $W^+ W^-$ and $Z Z$-decay products.
For a simple generic model, with SM-couplings to fermions, the 
parameters describing a CP violation in the higgs couplings to vector bosons 
can be determined to about 10\%.
If in addition the invariant-mass distribution is used to constrain model
parameters, 1--2 \% statistical uncertainty in the determination 
of the Higgs-boson couplings can be achieved.

\subsection*{Acknowledgements}

We are especially grateful to David Miller for stimulating discussions.
We would also like to thank  other colleagues from the 
ECFA/DESY Study for useful comments and suggestions.
M.K.~acknowledges partial
support by Polish Committee for Scientific Research, Grant 
5 P03B 121 20 (2003), and by the European Community's
Human Potential Programme under contract HPRN-CT-2000-00149 Physics
at Colliders.



\end{document}